\documentclass[conference]{IEEEtran}
\usepackage{flushend}
\usepackage{xcolor}
\usepackage{graphicx}
\usepackage{subfigure}
\usepackage{array}
\usepackage{booktabs}
\usepackage{booktabs}
\usepackage{graphicx}  % For resizing tables
\usepackage{pdflscape}

\usepackage{tcolorbox}
\usepackage{caption}
\usepackage{listings}
\usepackage{amsmath} % For math symbols

\usepackage{listings}
\usepackage{xcolor}
\usepackage{caption}
\usepackage{multirow}
\usepackage{makecell}
\usepackage{array}
\usepackage{graphicx}
\definecolor{MyDarkGreen}{cmyk}{1,0,1,0.5}

\AtBeginDocument{
  
}

\usepackage{graphicx}
\usepackage{subcaption}
\usepackage{listings}
\usepackage{xcolor}
\usepackage{booktabs}
\usepackage{graphicx}  % For resizing tables
\usepackage{verbatim}

\newcommand{\toolname}[0]{}
\renewcommand{\toolname}[0]{\emph{IntentContinuum}}

\ifCLASSINFOpdf
\else
\fi
\newcounter{insightcounter} % Create a counter for insights
\newcommand{\insight}[1]{
    \refstepcounter{insightcounter} % Increment and label the counter
    \noindent\textbf{\textit{Insight \theinsightcounter:}} \textit{#1} % Format the insight
}

\begin{document}
%
% paper title
% Titles are generally capitalized except for words such as a, an, and, as,
% at, but, by, for, in, nor, of, on, or, the, to and up, which are usually
% not capitalized unless they are the first or last word of the title.
% Linebreaks \\ can be used within to get better formatting as desired.
% Do not put math or special symbols in the title.
\title{\textit{IntentContinuum}: Using LLMs to Support Intent-Based Computing Across the Compute Continuum}

% author names and affiliations
% use a multiple column layout for up to three different
% affiliations 
%%%%%%%%%%%%%%%%%%%%%%%%%%%%%%%%%%%%%%%%%%%%%%%%%%
\author{
\IEEEauthorblockN{Negin Akbari\IEEEauthorrefmark{1}, John Grundy\IEEEauthorrefmark{1}, Aamir Cheema\IEEEauthorrefmark{1}, Adel N. Toosi\IEEEauthorrefmark{2}}
\IEEEauthorblockA{\IEEEauthorrefmark{1}Department of Software Systems and Cybersecurity, Monash University, Australia, \\ {\{negin.akbari, john.grundy, aamir.cheema\}}@monash.edu}
\IEEEauthorblockA{\IEEEauthorrefmark{2}School of Computing and Information Systems, The University of Melbourne, Australia, adel.toosi@unimelb.edu.au}
}
%%%%%%%%%%%%%%%%%%%%%%%%%%%%%%%%%%%%%%%%%%%%%%%%%%
% conference papers do not typically use \thanks and this command
% is locked out in conference mode. If really needed, such as for
% the acknowledgment of grants, issue a \IEEEoverridecommandlockouts
% after \documentclass

% for over three affiliations, or if they all won't fit within the width
% of the page, use this alternative format:
% 
%\author{\IEEEauthorblockN{Michael Shell\IEEEauthorrefmark{1},
%Homer Simpson\IEEEauthorrefmark{2},
%James Kirk\IEEEauthorrefmark{3}, 
%Montgomery Scott\IEEEauthorrefmark{3} and
%Eldon Tyrell\IEEEauthorrefmark{4}}
%\IEEEauthorblockA{\IEEEauthorrefmark{1}School of Electrical and Computer Engineering\\
%Georgia Institute of Technology,
%Atlanta, Georgia 30332--0250\\ Email: see http://www.michaelshell.org/contact.html}
%\IEEEauthorblockA{\IEEEauthorrefmark{2}Twentieth Century Fox, Springfield, USA\\
%Email: homer@thesimpsons.com}
%\IEEEauthorblockA{\IEEEauthorrefmark{3}Starfleet Academy, San Francisco, California 96678-2391\\
%Telephone: (800) 555--1212, Fax: (888) 555--1212}
%\IEEEauthorblockA{\IEEEauthorrefmark{4}Tyrell Inc., 123 Replicant Street, Los Angeles, California 90210--4321}}

% use for special paper notices
%\IEEEspecialpapernotice{(Invited Paper)}

% make the title area
\maketitle

% As a general rule, do not put math, special symbols or citations
% in the abstract
%\textcolor{purple}{FYI- Page limits: Every “Regular Paper” manuscript can include 7 to 10 pages for the main contents (including all text, footnotes, figures, tables and appendices) with additional pages for appropriate references.((We need to reduce a bit since we have one paragraph and a table in page 11- Please let me know which section we can reduce.)) }

\begin{abstract}
The increasing proliferation of IoT devices and AI applications has created a demand for scalable and efficient computing solutions, particularly for applications requiring real-time processing. The \emph{compute continuum} integrates edge and cloud resources to meet this need, balancing the low-latency demands of the edge with the high computational power of the cloud. However, managing resources in such a distributed environment presents challenges due to the diversity and complexity of these systems. Traditional resource management methods, often relying on heuristic algorithms, struggle to manage the increasing complexity, scale, and dynamics of these systems, as well as adapt to dynamic workloads and changing network conditions. Moreover, designing such approaches is often time-intensive and highly tailored to specific applications, demanding deep expertise. In this paper, we introduce a novel framework for intent-driven resource management in the compute continuum, using large language models (LLMs) to help automate decision-making processes. Our framework ensures that user-defined intents -- such as achieving the required response times for time-critical applications -- are consistently fulfilled. In the event of an intent violation, our system performs root cause analysis by examining system data to identify and address issues. This approach reduces the need for human intervention and enhances system reliability, offering a more dynamic and efficient solution for resource management in distributed environments.

\textit{Keywords- Compute Continuum, Resource Management, Intent-driven scheduling, LLM.}
\end{abstract}

% no keywords

% For peer review papers, you can put extra information on the cover
% page as needed:
% \ifCLASSOPTIONpeerreview
% \begin{center} \bfseries EDICS Category: 3-BBND \end{center}
% \fi
%
% For peerreview papers, this IEEEtran command inserts a page break and
% creates the second title. It will be ignored for other modes.
\IEEEpeerreviewmaketitle

\section{Introduction} 
The rapid growth of the AI-driven Internet of Things (IoT) has led to a massive increase in smart devices, each generating large amounts of data \cite{lee2015internet}. This surge requires scalable storage and processing solutions, such as cloud computing. However, cloud computing alone may not be suitable for applications that need real-time processing or strong privacy protections~\cite{tabrizchi2020survey}. This challenge has led to the development of the \emph{``compute continuum''} that combines edge and cloud resources to ensure smooth and efficient operations across a wide range of applications \cite{russo2023serverless}. In this integrated system, edge devices often have limited computing power and storage, yet they must handle tasks that require low latency and quick response times \cite{9083958}. On the other hand, cloud resources offer greater computational power and storage but involve higher latency and potential privacy concerns \cite{sajid2013cloud}. Resource management is thus particularly challenging due to the need to balance the diverse requirements of both edge and cloud environments \cite{danelutto2024structuring}.  Efficient resource management must account for these differences, ensuring that tasks are allocated optimally between the edge and cloud. This involves dynamically adjusting to changes in workload, network conditions, and energy consumption, all while maintaining the seamless operation of applications. 

The compute continuum, with its diverse and distributed nature, poses significant challenges for resource management and scheduling, which are traditionally categorized as NP-hard in its general case %\textcolor{red}{(CCGRID comments:  this statement holds when the problem formulation involves integer variables. However, problems with linear formulations may not fall into this category. So the statement should be rewritten to clarify the scope. *** see my comment)} 
\cite{guzek2015survey}. Solving these problems optimally becomes computationally infeasible for large instances due to the exponential growth of possible solutions. To address this, heuristic \cite{sangaiah2020iot} and meta-heuristic approaches \cite{sharma2022systematic}, such as genetic algorithms (GA) \cite{katoch2021review} and simulated annealing \cite{guilmeau2021simulated}, are widely used to find near-optimal solutions within a reasonable timeframe. However, designing efficient heuristic algorithms often requires domain expertise and problem-specific tuning, which limits their adaptability to the heterogeneous and dynamic demands of the compute continuum \cite{danelutto2024structuring}.
As a result, current efforts focus on automating these processes to reduce dependency on human intervention and specialized algorithms tailored to specific problems. 

With the rapid advancements in artificial intelligence (AI) and the increasing prevalence of machine learning (ML) techniques, generative AI and large language models (LLMs) have achieved a level of intelligence comparable to that of human experts. In this paper, we explore whether ``\textit{LLMs can complement—or even replace—both human experts and the need for developing specialized algorithms, offering a more generalized and efficient approach for resource management in compute continuum}''. We aim to leverage the capabilities of general purpose LLMs, such as ChatGPT, to develop novel intent-driven resource management techniques for distributed systems. We believe that LLMs, with their ability to process and analyze vast amounts of data, offer a promising solution to these challenges while simplifying the overall process. By leveraging the adaptive and contextual understanding capabilities of LLMs, resource management can be more dynamic and responsive to the continuum's needs.

Thus, we present a novel framework, \toolname, designed to manage and optimize application deployments and operations across the continuum, ensuring that user-defined intents—particularly service-level objectives (SLOs) for response times in image-processing IoT applications—are consistently met. Our framework uniquely monitors, analyzes, and addresses any deviations from these intents, thereby maintaining optimal performance across the compute continuum. 

\toolname~uses the integration of LLMs, specifically Open AI GPT-4o, as a central decision-making entity within the framework. When a violation of the predefined intent occurs, our algorithm employs GPT-4o to conduct a comprehensive root cause analysis. By processing system data—including network topology, cluster information, and real-time monitoring metrics—GPT-4o determines whether the issue originates from computational constraints (such as CPU or memory shortages) or network-related problems (like bandwidth limitations or link congestion). Following the identification of the root cause, GPT-4o suggests specific actions to resolve the issue, drawing from a predefined list of potential solutions. These actions are automatically configured in the system, initiating a continuous feedback loop to ensure that the user-defined intent is always satisfied. This loop allows the system to adapt to changes in workload, network conditions, or other environmental factors, reducing the need for manual intervention and enhancing the reliability of the compute continuum environment. The \textbf{key contributions} of this work include:

\begin{itemize}
    \item We propose \toolname, an innovative monitoring and automated reconfiguration framework for the compute continuum, designed to support a range of user-specified intent-driven performance criteria;
    \item We propose an LLM-powered root cause analysis and automated reconfiguration platform for \toolname~to manage user intents in the compute continuum. To the best of our knowledge, we are among the first to leverage LLMs as resource managers in this way, with initial evaluations showing promising results;
    \item We develop a prototype of \toolname~using industry-standard techniques and release it as an open-source solution to demonstrate its effectiveness in real-world scenarios; and 
    \item We conduct an extensive evaluation of the \toolname~platform’s performance across various practical scenarios and user-defined intents, complemented by a preliminary feasibility study assessing its potential impact.
\end{itemize}

%The remainder of the paper is organized as follows:
The remainder of the paper is organized as follows. Section \ref{sec:motiv} presents our motivation, outlining the context and challenges that drive our approach. Section \ref{sec:overview} provides a comprehensive description of the system architecture, including its key components and architecture. In Section \ref{sec:performance}, we conduct a detailed performance evaluation, discussing the metrics and results in depth. Section \ref{sec:discuss} delves into the discussion and addresses potential limitations of our work. Section \ref{sec:related} discusses related work, comparing our approach to existing research in the field. Finally, the paper concludes in Section \ref{sec:sumup}, summarizing our findings and suggesting potential avenues for future research.

\section{Motivation}\label{sec:motiv}

Consider a real-time image-processing application for predictive maintenance in an Industry 4.0 setting, where strategically placed camera sensors capture data from production lines and equipment. These sensors work in tandem with edge and cloud components to analyze images and detect potential issues. Such an application must meet strict SLOs, like low response times for detecting faults, despite dynamic conditions such as fluctuating network latency or intermittent connectivity.

Ensuring these SLOs are consistently satisfied is a challenging and non-trivial task. The factory environment relies on a wide variety of computational resources across the edge-cloud continuum where the application is deployed, coupled with evolving network conditions and shifting operational priorities—for example, prioritizing real-time issue detection during peak hours versus energy efficiency during off-hours. Developing robust resource management techniques to dynamically adapt and ensure these SLOs are met is essential, as traditional rule-based approaches often fall short in handling such variability and unpredictability.

Thus, we aim to address the following key research questions:

RQ1: --``\textit{Can a large language model (LLM), such as GPT-4o, enhance real-time resource management in the compute continuum for IoT applications?}''

RQ2: --``\textit{How effective is such an LLM-powered approach in identifying and resolving issues when performance goals (e.g., response times) are violated in the compute continuum?}''

\section{Our Approach} \label{sec:overview}

\subsection{IntentContinuum Architecture}

\begin{figure*}[tph!t]
\centering
\includegraphics[width=0.8\textwidth]{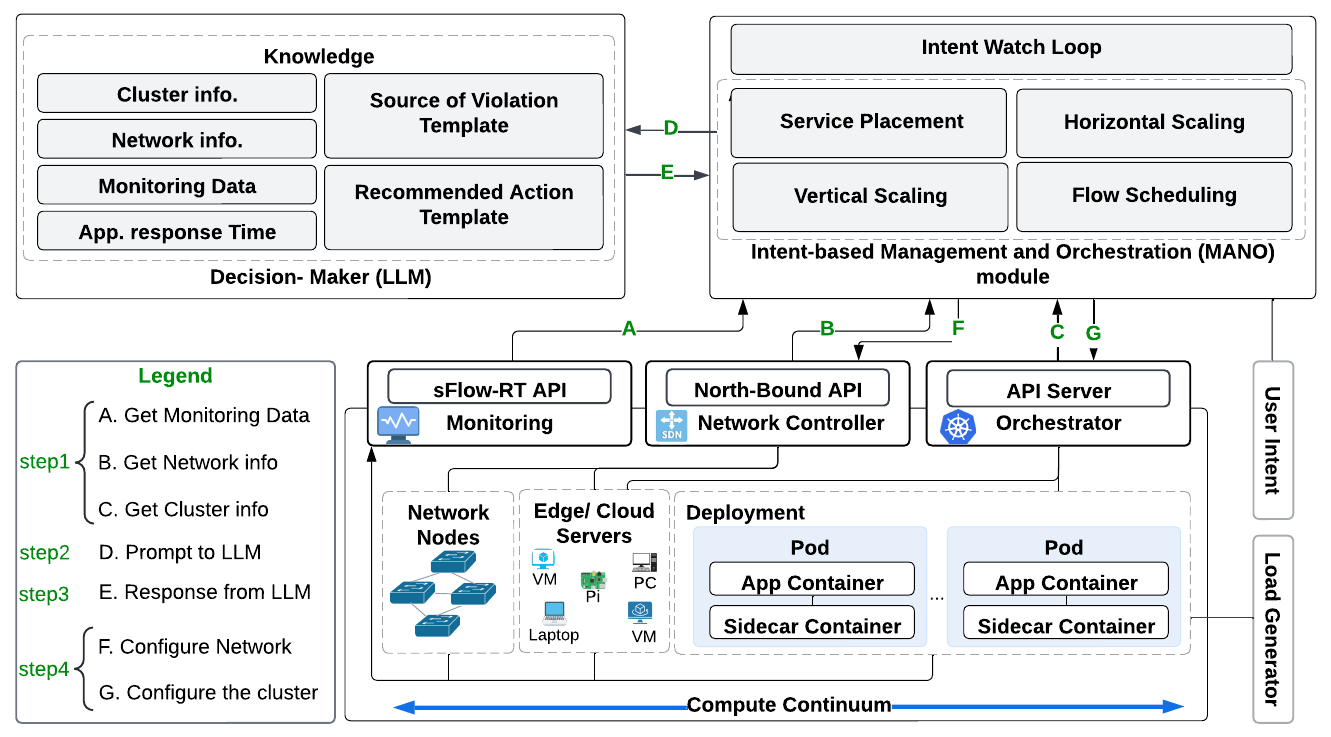}
\caption{Architecture of our \toolname~system}
\label{Fig:arc}
\vspace{-0.5cm}
\end{figure*}

Figure \ref{Fig:arc} illustrates the architecture of our \toolname~framework, designed to manage and optimize application deployment and operations within a compute continuum environment. The source code for the framework is available on the anonymous GitHub repository.\footnote{https://github.com/Git-anonymous-creator/anonymous} %\textcolor{red}%{(CCGRID comment: The code should have been provided to the reviewers through an anonymized repository.*** see my comment)} 
Below we describe its key components and their roles within the system.

\noindent \textbf{Target Compute Continuum Environment:}
At the bottom of Figure \ref{Fig:arc}, we illustrate the target compute continuum environment, where nodes, including both edge and cloud servers, are managed by a container orchestrator such as Kubernetes. Various microservices (pods in Kubernetes) run on these nodes, interconnected through network switches controlled by a Software Defined Network (SDN) controller. Each element in this continuum, including edge and cloud servers, network switches, and application pods—utilizing sidecar containers\footnote{https://kubernetes.io/docs/concepts/workloads/pods/sidecar-containers/}—is continuously monitored by our monitoring tool. %To simulate traffic within the application, we use Locust as the load generator to send HTTP requests to a chain of microservices, representing our IoT application.
Requests from sensors or end users are directed to the application's ingress gateway for processing. Users can define specific SLOs for their application, such as response time targets or energy efficiency, based on high-level intents. In this context, intents represent high-level user goals, such as minimizing latency or optimizing energy consumption, while SLOs define measurable performance targets, such as maintaining an average response time below a specified threshold. For the purposes of this paper, these two terms are used interchangeably.

%\textcolor{red}{(CCGRID comment: The term user-defined intent seems to be a redefinition of the user SLO. If there is any difference, it should be specified. Otherwise, why use a different term? *** see my comment)} 
The dynamic nature of this environment, as highlighted in our contributions, demands ongoing adaptation to meet these user-defined intents. 

\noindent \textbf{Intent-based Management and Orchestrator (MANO) Module:}
An \textit{Intent Watch Loop} continuously monitors the defined intent of the compute continuum target application for any violations. Upon detecting a violation, the watch loop triggers the \textit{Decision-Maker} module to resolve the issue. The \textit{Monitoring} module, in the target compute continuum, retrieves information on application performance, network status, and cluster details through concurrent API calls (step 1 in the figure). This data helps create a clear picture of the system's current state and identify deviations from the user-defined intents. 

The collected data is then sent to the \textit{Decision-Maker} module via an API call, where the root cause of the violation is diagnosed (step 2). Based on the analysis, appropriate corrective actions are recommended to resolve the issue and ensure compliance with the intents (step 3). Validated by recent research in cloud-edge computing \cite{millnert2020holoscale, marchese2022network, toosi2019elasticsfc,10643932}, we consider four corrective actions: 1) \textit{\underline{service placement}} (relocating pods to a more suitable node to optimize resource utilization and minimize latency), 2) \textit{\underline{horizontal scaling}} (adjusting the number of pod replicas to handle dynamic workloads efficiently), 3) \textit{\underline{vertical scaling}} (modifying CPU and memory allocations to enhance stability and resource efficiency), and 4) \textit{\underline{flow scheduling}} (rerouting application network traffic through optimal paths to mitigate congestion and network instability). By integrating these approaches, our method ensures improved application responsiveness and resource allocation in the compute continuum. Finally, the Network Controller and Orchestrator apply the necessary reconfigurations on the relevant nodes via their APIs (step 4) to address the root causes of intent violation.

\begin{figure*}[t]%[width=1\textwidth]
    \centering
    \subfigure[API Calls]{\label{Fig:api}
          \includegraphics[width=0.322\linewidth]{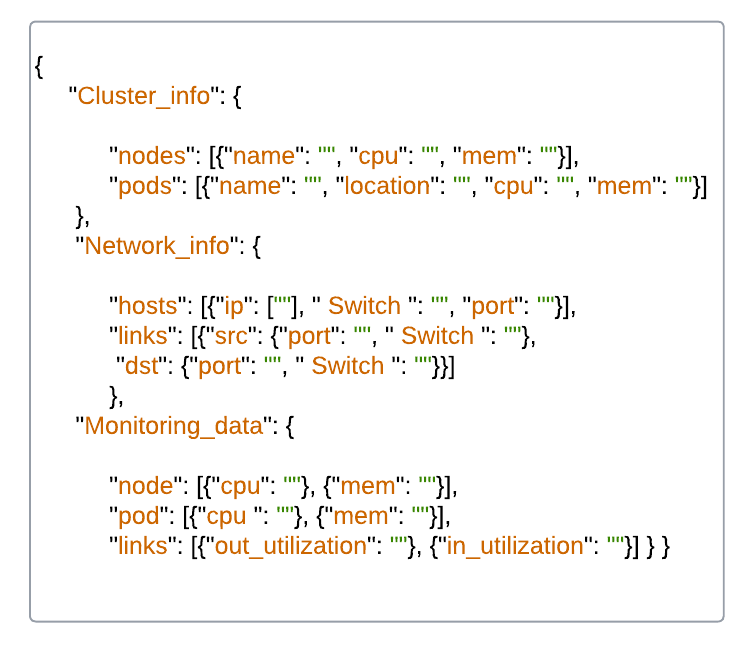}}    
    \subfigure[Source of Violation Template]{\label{Fig:srctemplate}
          \includegraphics[width=0.32\linewidth]{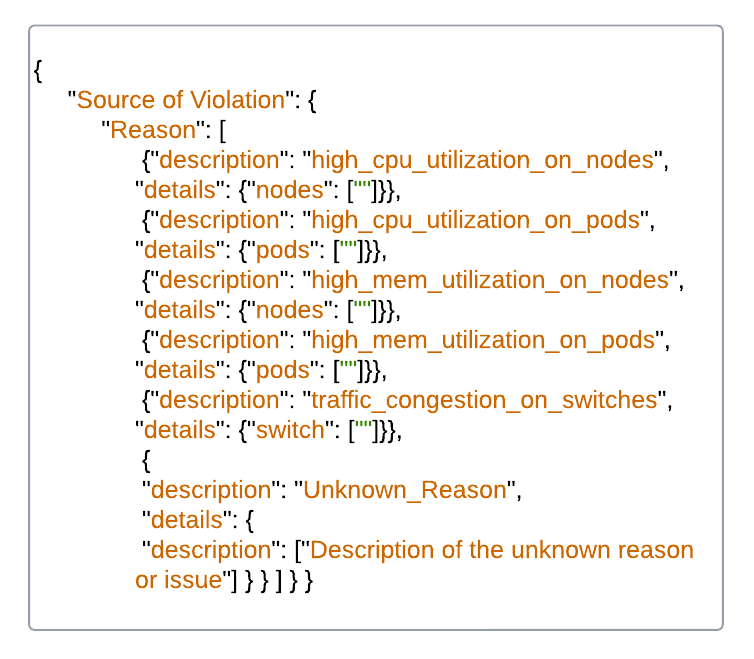}}
    \subfigure[Recommended action Template]{\label{Fig:action_template}
          \includegraphics[width=0.32\linewidth]{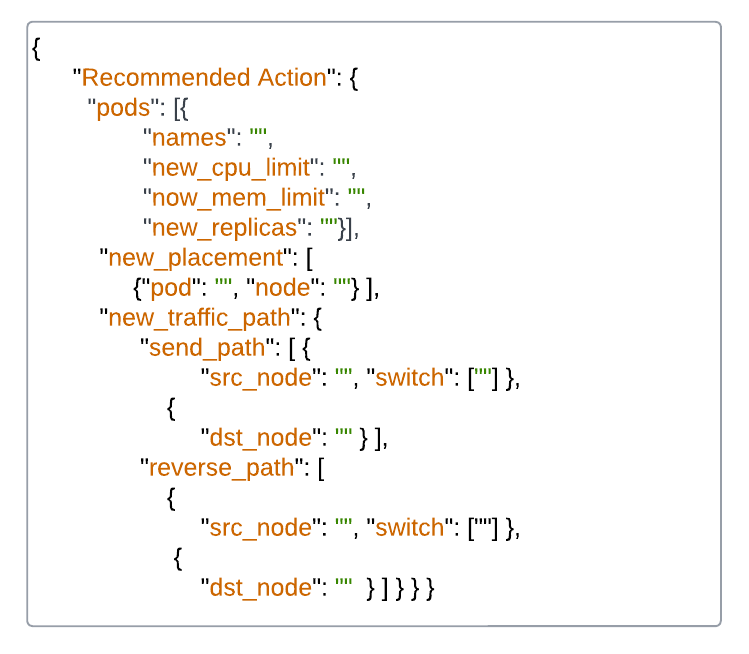}}    
    \caption{Examples of JSON objects sent to the LLM in order to provide structured inputs and outputs %\textcolor{red}{\\ -- Negin: use correct json format for these figures}
    }
    \label{fig:training}
    \vspace{-0.4cm}
\end{figure*}

\noindent \textbf{Decision-Maker:}
The core of the system is the \textit{Decision-Maker} module, powered by a large language model (LLM), specifically OpenAI GPT-4o in this work. This module processes data collected from various system components and orchestrates responses to maintain system performance. The \textit{Decision-Maker} begins by receiving a comprehensive narrative that describes the system's architecture and operational dynamics,  providing essential context for GPT-4o. The collected data is formatted as JSON objects (Figure \ref{Fig:api}), which encapsulate three primary categories: (1) \textit{\underline{Cluster information}}, detailing nodes and pods along with their CPU and memory allocations; (2) \textit{\underline{Network information}}, representing hosts, switches, ports, and interconnecting links; and (3) \textit{\underline{Monitoring data}}, which tracks real-time resource utilization at the node, pod, and network link levels. These structured inputs ensure a comprehensive and machine-readable representation of the system’s state, enabling accurate root cause analysis. When a deviation from expected performance is detected (indicating an intent violation), the \textit{Decision-Maker} identifies the root cause by analyzing the collected data using a structured \textit{Source of Violation Template} (Figure \ref{Fig:srctemplate}). These sources of violations are the major performance challenges reported in cloud and edge computing \cite{fareghzadeh2018dynamic, toosi2019elasticsfc}. Following this analysis, the module leverages a \textit{Recommended Action Template} (Figure \ref{Fig:action_template}) to propose targeted corrective actions, addressing issues related to computational resources, network conditions, or other operational factors. To enhance GPT-4o’s decision-making, the prompt is structured using a template-based approach and few-shot learning \cite{brown2020language}. Few-shot learning is incorporated directly within the prompt, providing structured examples of intent violations, root cause analyses, and corrective actions. This approach ensures consistent and optimized decision-making without modifying the underlying model. The \textit{Decision-Maker} integrates its recommendations with the \textit{MANO} module, enabling seamless implementation of corrective measures. 

\begin{figure*}[tph!t]
\centering
\includegraphics[width=0.89\textwidth]{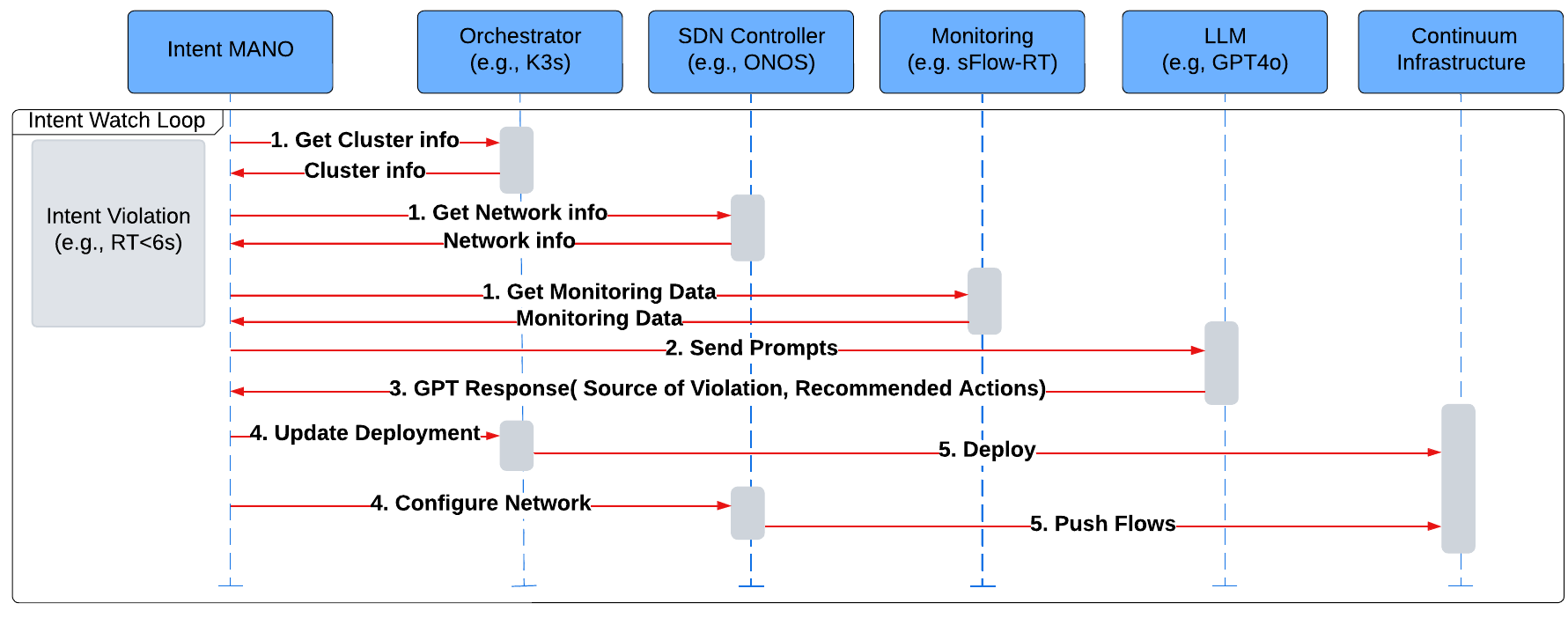}
\caption{Sequence Diagram of key \toolname~process steps}
\label{Fig:seq}
\vspace{-0.5cm}
\end{figure*}

\subsection{Detailed Decision Making Process}

Figure \ref{Fig:seq} illustrates the sequence diagram of our \toolname~tool decision making process, highlighting the interactions between each module when a defined intent is violated. The process begins with the \textit{Intent Watch Loop}, a continuous monitoring system that detects intent violations, such as when the application response times exceed a specific threshold (e.g., smaller than 6 seconds). Upon detecting a violation, the system initiates corrective actions.

In the first stage, \textit{MANO} makes an API call to the \textit{Orchestrator} to gather cluster information, including the state and configuration details of the applications and services running within the cluster. Concurrently, an API call is made to the \textit{SDN Controller} to obtain network information, detailing the status and configuration of network resources. Another concurrent API call is directed to the \textit{Monitoring} module to collect comprehensive monitoring data, such as CPU and memory utilization of nodes and pods in the cluster, as well as the amount of traffic on each switch interface. These API calls are collectively labeled as 1 in the sequence diagram for reference. Given the large volume of monitoring data, we condense the information before sending it to the \textit{LLM} by aggregating average metrics over multiple segments: `pre-violation' and `violation.' These segments are discussed in more detail later.

\noindent \textbf{Response Time SLO}: To maintain the response time SLO within the specified range and prevent the system from overreacting to temporary fluctuations or noise, we calculate the \emph{Exponential Moving Average (EMA)} of response time, which is represented as:
\[
\text{EMA\_RT}_t = (1 - \alpha) \times \text{EMA\_RT}_{t-1} + \alpha \times \text{RT}_t
\tag{1} \label{eq:estimated_rt}
\]

Here, $RT_t$ denotes the response time of the most recent request at time $t$, and $EMA\_RT_t$ represents the exponential moving average of the response time at time $t$. In this formula, \( \alpha\) is the smoothing factor, for example, \( \alpha = 0.02\). This method continuously updates $EMA\_RT$, assigning more weight to recent samples. A violation is detected when the EMA\_RT exceeds or drops below predefined thresholds, marking a significant performance deviation. As soon as a violation is detected, the collected details are sent to \textit{LLM} for analysis and find the appropriate actions to rectify the issue. This process is labeled as 2 in the sequence diagram for clarity.

For the `pre-violation' segment, we analyze the system's response times across windows, e.g., last 30 requests. The pre-violation data is collected using a \textbf{Fixed-Time-Window Aggregation} method, meaning data is aggregated into fixed-length windows immediately preceding the violation. Mathematically, this is represented as:
\[
\text{Pre-Violation} = \{..., \omega_{t-2}, \omega_{t-1}\}
\]
where \( \omega_{t-1} \), \( \omega_{t-2} \), etc., represent the windows immediately preceding the violation, while the window \( \omega_t \) contains the violation and represents the anomaly. This approach helps identify relevant system performance metrics at the time of the response time breach, providing insights into potential causes. By structuring the data this way, we provide \textit{LLM} with a concise yet informative view of the system’s performance both before and during the violation for further analysis.

Next, \toolname~sends prompts to the \textit{LLM}, step 2 in the sequence diagram, which processes the collected data. The \textit{LLM} analyzes the situation, identifies the source of the violation, and recommends corrective actions (step 3). Following the \textit{LLM}'s recommendations, the \textit{Orchestrator} updates the deployment, step 4 and step 5, by making the necessary adjustments to the applications or configurations. The \textit{SDN Controller} also reconfigures the network to align with the updated deployment, step 4 and step 5. This way, our framework enables a dynamic, automated reactions to intent violations, ensuring system resilience and optimal performance.

\section{Performance Evaluation}\label{sec:performance}

In this section, we present a performance evaluation of \toolname~using an example image processing application as a representative scenario. We describe our experimental testbed setup in detail, followed by an in-depth analysis of the evaluation results. These results provide valuable insights into the performance and capabilities of our proposed \toolname~method using LLMs. We benchmark our approach against existing methods to highlight its effectiveness,  advantages and limitations. Our results provide valuable insights into the performance and capabilities of our proposed \toolname~method using LLMs.

\subsection{Experimental Setup}\label{sec:setup}
\begin{figure}[t]
\centering
\includegraphics[width=0.79\columnwidth]{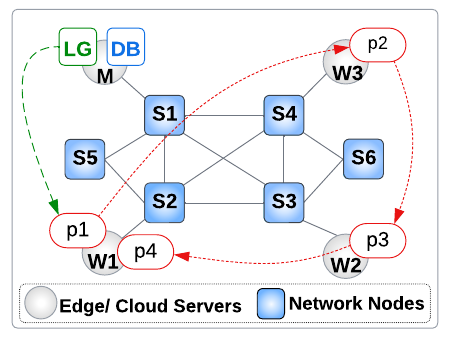}
\caption{Scenario's Network Topology}
\label{Fig:topo}
\vspace{-0.6cm}
\end{figure}

To evaluate \toolname, we leverage \textit{iContinuum} emulator~\cite{10643932}, 
creating a controlled environment for studying the proposed intent-driven resource management. 
\textit{iContinuum} offers a flexible platform for deploying real-world applications over the edge-cloud continuum and simulates practical conditions by integrating SDN controllers, Mininet-based network emulation, and containerized applications managed by Kubernetes.
Figure \ref{Fig:topo} illustrates the topology of the experimental setup. We used a Kubernetes cluster consists of one Master node (M) and three Worker nodes (W1–W3), interconnected via Open vSwitches (OVS),\textit{ S1–S6} in the figure, which are managed by ONOS as the SDN controller. The application deployed on the Kubernetes cluster is an image processing application composed of a chain of four microservices, hosted on four pods (p1, p2, p3, and p4). These pods are distributed across the cluster nodes and communicate through the red dashed arrows in the figure, which represent the flow of data. Additionally, the Master node hosts a Load Generator (LG) that sends traffic to the pods and a Database (DB) that records pod activity. The green dashed arrow shows HTTP requests sent from Locust, acting as the Load Generator. Locust sends a 499.69 KB image to the entry microservice in the chain, hosted on pod p1. Detailed configuration of the nodes are presented in Table \ref{tab:vm}.

\begin{table}[h]
    \centering
    \begin{tabular}{|c|c|c|c|c|}
        \hline
        \textbf{Node} & \textbf{OS} & \textbf{Arch.} & \textbf{RAM} & \textbf{vCPU} \\
        \hline
        \ Edge Servers & Ubuntu 20.04 LTS & amd64 & 64GB & 32 \\
        \hline
        \ SDN Controller & Ubuntu 20.04 LTS & amd64 & 64GB & 32 \\
        \hline
    \end{tabular}
    \captionsetup{justification=centering}
    \caption{Configuration of Nodes}
    \label{tab:vm}
    \vspace{-0.3cm}
\end{table}

We define the intent as a response time that must remain within a specified range, bounded by upper and lower thresholds, to ensure efficient resource utilization and consistent application performance.
An intent violation is detected based on EMA\_RT (as defined in Eq.~\eqref{eq:estimated_rt}) when it either exceeds the upper threshold or drops below the lower threshold.
For our experiments, we set the upper threshold to 3 seconds and the lower threshold to 1 second, determined through initial testing under varying traffic conditions. These thresholds represent a reasonable range for our application's operational tolerance while ensuring stability and responsiveness. This dual-threshold approach ensures a balance between cost-efficiency and performance. The upper threshold safeguards against performance degradation by flagging insufficient resource allocation, while the lower threshold prevents resource wastage by identifying over-provisioning. If the EMA\_RT exceeds the upper threshold, the system flags a violation and sends details to GPT, which identifies the root cause and recommends corrective actions. Similarly, if the EMA\_RT drops below the lower threshold, the system consults GPT to analyze the situation and recommend optimization strategies to ensure efficient resource usage.

To reflect the varying computational demands of each pod, resource limits are configured for the pods as presented in Table \ref{tab:podslimit}. Since pod p3 has higher processing requirements compared to the other pods, it is assigned increased resource limits. Pod placement is initially managed by Kubernetes' default scheduler; however, pod p1 is fixed on Worker1 to serve as the first entry point of the application, ensuring consistent traffic flow into the system. The placement of the pods within the cluster is illustrated in Figure \ref{Fig:topo}. The pods are managed behind Kubernetes services, which provides load balancing and facilitates communication between the pods and external systems. Additionally, the system is designed to support multiple replicas of the same pod to handle varying workloads and ensure high availability. 

\begin{table}[ht!b]
    \centering
    \begin{tabular}{|c|c|c|}
        \hline
        \textbf{Pod name} & \textbf{CPU limit (core)} & \textbf{Mem limit (MiB)} \\
        \hline
        p1, p2, p4 & 0.3 & 312 \\
        \hline
        p3 & 0.5 & 512 \\
        \hline
    \end{tabular}
    \captionsetup{justification=centering}
    \caption{Configuration of Pods}
    \label{tab:podslimit}
    \vspace{-0.3cm}
\end{table}

Traffic generation is managed using Locust for a duration of 900 seconds, simulating user loads with the following sequence: [10, 20, 15, 10, 5, 20, 10] users, with a spawn rate of 1 user per second. The interval between each load change is set to 120 seconds. During the experiment, if the system detects that the EMA\_RT exceeds the maximum threshold or falls below the minimum threshold, it flags the violation and sends an API call to GPT to address the violation. After implementing these actions, the system introduces a ``waiting-time" metric, defined as the duration during which the system halts monitoring and evaluation to allow the network to stabilize. For our setup, the ``waiting time" is set to 1 minute after each corrective action, determined through experimental testing to ensure the system reaches a stable state before further evaluation and reconfiguration.

\subsection{Results and Analysis}\label{sec:analysis}
In this section, we present experiments demonstrating how \toolname~effectively satisfies the intent by addressing both computing and networking actions.
\noindent \subsubsection{\textbf{Computing Experiment}}
\begin{figure}[t]
\centering
\includegraphics[width=0.95\columnwidth]{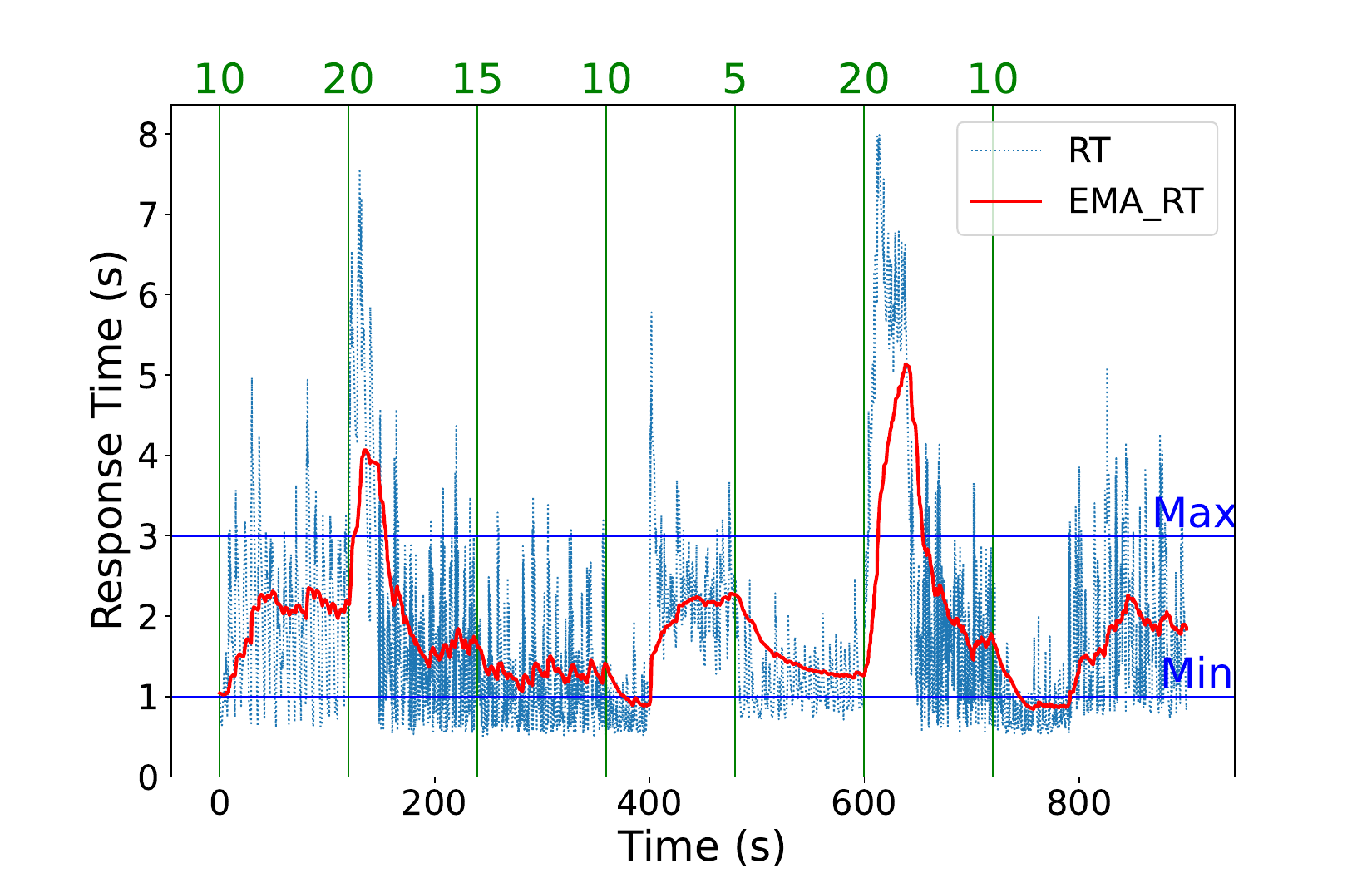}
\caption{Application Response Time for Computing Experiment}
\label{Fig:rtintent}
\vspace{-0.6cm}
\end{figure}

Figure \ref{Fig:rtintent} presents the application response time under varying load levels generated by Locust. The x-axis represents the time (in seconds) during the experiment, while the y-axis indicates the response time (in seconds) for both RT, indicating the response time for each individual request (dashed blue line), and EMA\_RT (red line), calculated using the formula described in Eq.~\eqref{eq:estimated_rt}. The graph includes the upper (max) and lower (min) thresholds, set at 3 seconds and 1 second, respectively, offering a clear depiction of the application's performance relative to the defined intent boundaries. The green numbers displayed at the top of the graph represent the number of active users sending requests (system load).

At the start of the experiment, with 10 users, EMA\_RT stays within the defined thresholds, as shown in the graph. After 120 seconds, the user count increases to 20, causing a sharp spike in EMA\_RT that exceeds the upper threshold. At this point, \toolname~promptly identifies the source of the violation and recommends corrective actions, successfully restoring the response time to within acceptable limits after about 30 seconds. A one-minute waiting period follows the corrective measures to ensure system stability. At 240 seconds, the user count decreases to 15, leading to a drop in EMA\_RT. No violations occur during this period, as the previous corrective actions proved highly effective in maintaining system stability. Subsequently, at 360 seconds, the user count decreases further to 10, resulting in minor intent violations around the time of 380 seconds. \toolname~quickly addresses this violation, ensuring the response time returns to within acceptable boundaries. When the user count drops to 5 at 480 seconds, the system maintains the desired response time without any violations. The load then increases again to 20 users at 600 seconds, causing another violation of the maximum threshold. \toolname~resolves this issue effectively within about 40 seconds. At 720 seconds, when the user count decreases to 10, EMA\_RT briefly dips below the lower threshold. Once more, \toolname~swiftly addresses the deviation, restoring the response time to the defined SLO within around 40 seconds.

\begin{figure*}[t]%[width=1\textwidth]
    \centering
    \scriptsize
    \subfigure[IntentContinuum]{\label{Fig:lifespan}
          \includegraphics[width=0.23\linewidth]{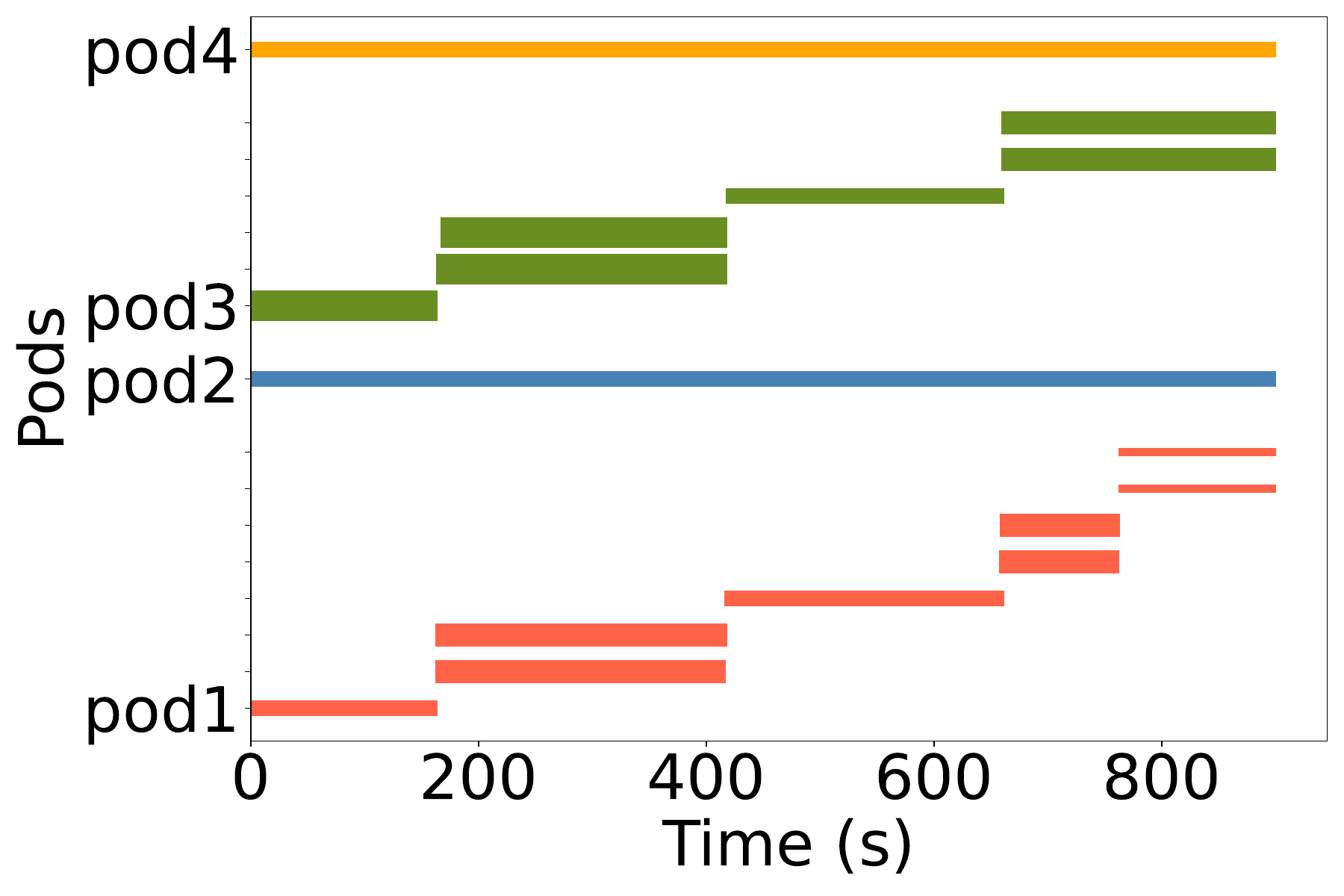}}
    \subfigure[CPU threshold- 70\%]{\label{fig:70}
          \includegraphics[width=0.23\linewidth]{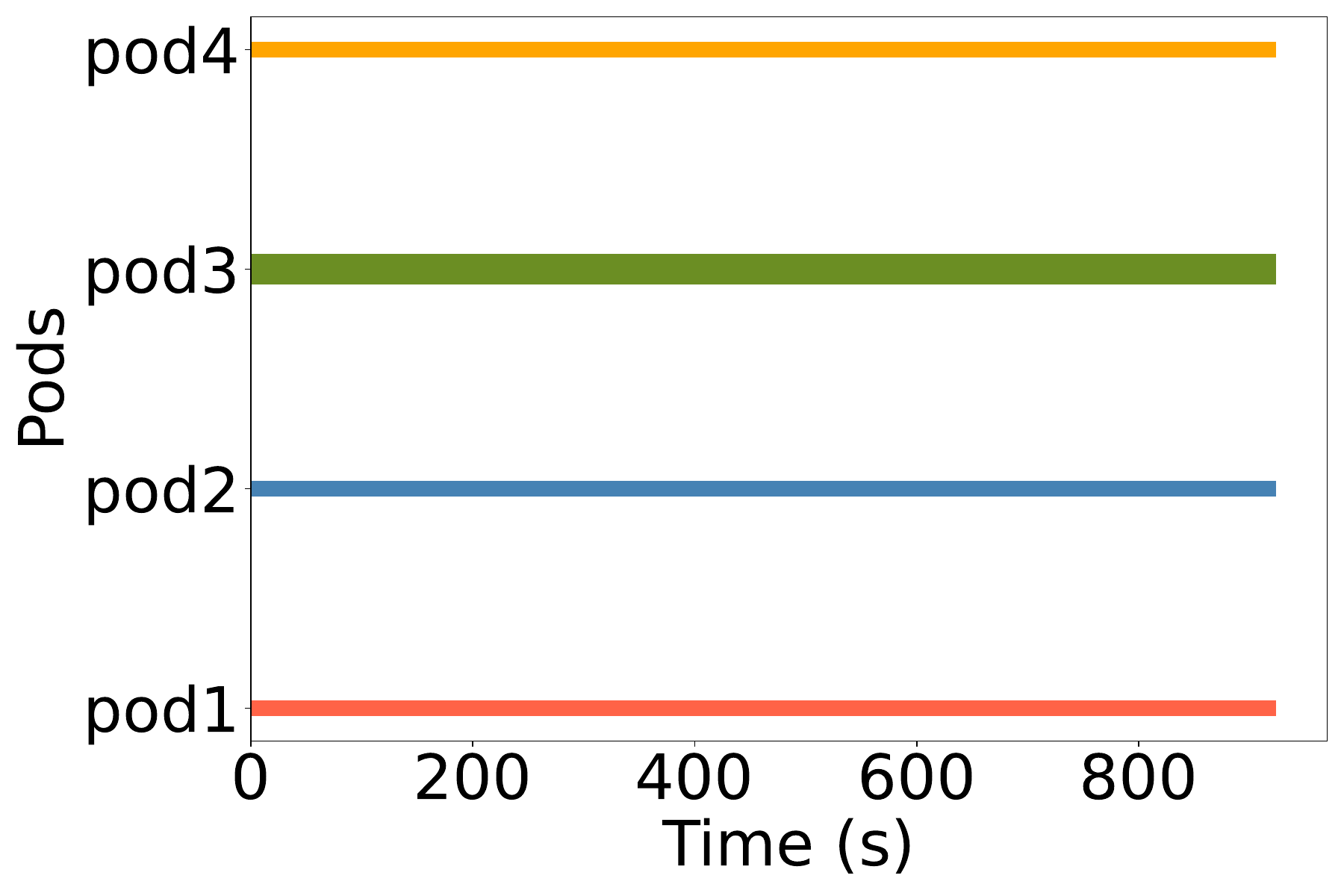}}    
    \subfigure[CPU threshold- 60\%]{\label{fig:60}
          \includegraphics[width=0.23\linewidth]{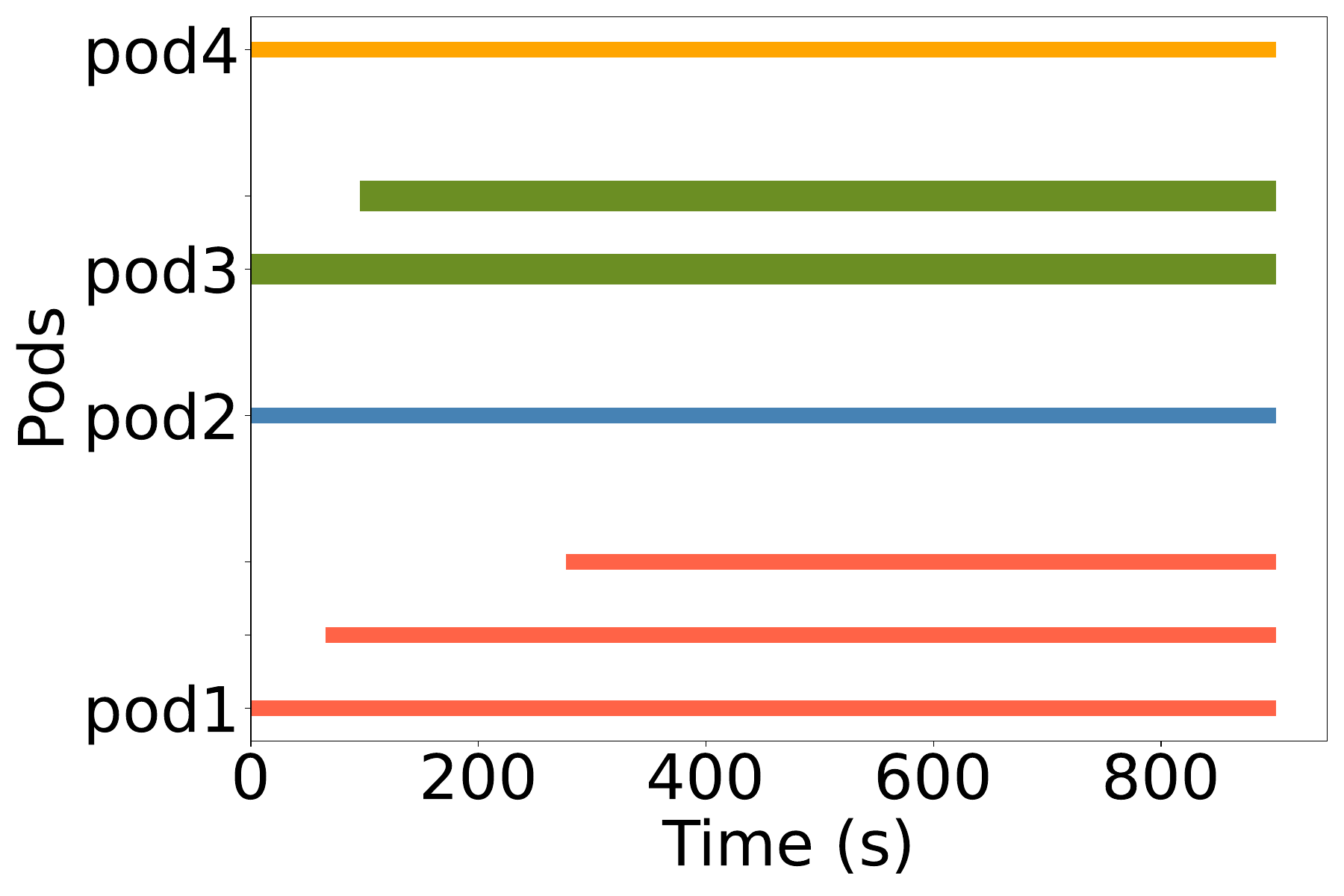}}
    \subfigure[CPU threshold- 50\%]{\label{fig:50}
          \includegraphics[width=0.23\linewidth]{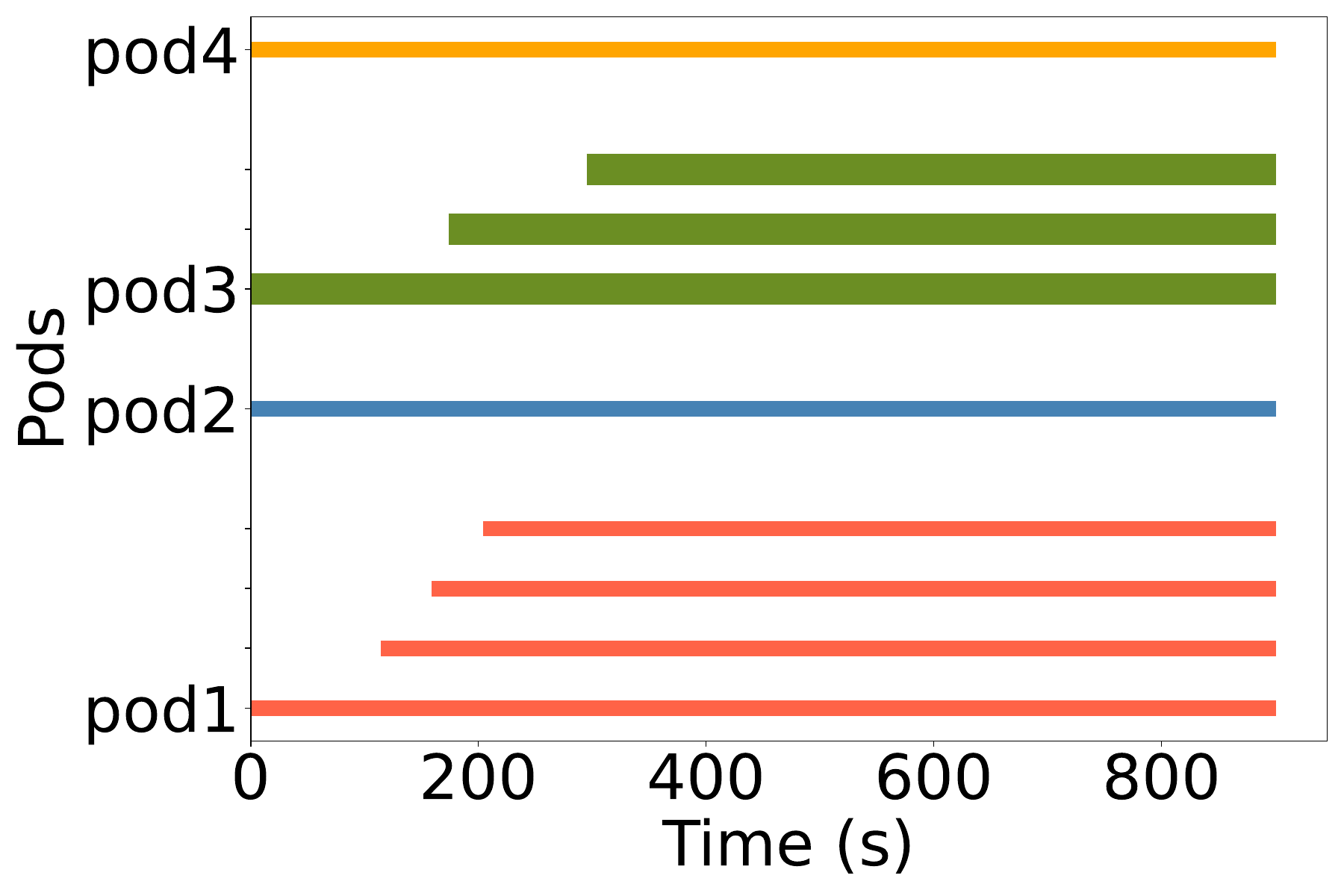}}    

    \caption{Pods Lifespans and The Number of Replicas Across Methods- Line Thickness Represents Vertical Scaling, While the Number of Lines Indicates Horizontal Scaling}
    \label{fig:all-pod-age}
    \vspace{-0.4cm}
\end{figure*}

\insight{Results demonstrate that \toolname, leveraging LLM recommendations, can dynamically adapt to varying load levels while maintaining the application's response time within the defined thresholds.}

To resolve these violations, \toolname~employs a range of actions, including adjusting pod placement, scaling replicas up or down, and dynamically reallocating resources such as CPU and memory. This demonstrates the system's ability to apply a variety of strategies—or a combination of them—to address violations effectively.  In the following we discuss, the detailed actions recommended by \toolname.

Figure \ref{Fig:lifespan} shows how the number (horizontal scaling) and size (vertical scaling) of replicas were adjusted during the experiment. Multiple lines represent horizontal scaling, while line thickness indicates CPU allocation per replica (0.5 cores for the thickest to 0.2 cores for the thinnest). Pod1, handling incoming traffic, and pod3, with heavier processing, scale dynamically, while pods 2 and 4 remain stable. 

Table \ref{tab:podstate} illustrates how pod placement changes during the experiment due to detected violations and the corresponding corrective actions. It represents different pod placements: \textit{init} denotes the initial placement, while V1, V2, V3, and V4 correspond to the system's responses to the first, second, third, and fourth violations, respectively. The table also indicates how pods are deployed on each node at each stage. The system recommends relocating pod3 in response to the first (V1) and third (V3) violations, while the other pods remain on their initial placement throughout the experiment.

\begin{table}[ht!b]
    \centering
    \begin{tabular}{|c|c|c|c|}
        \hline
        \textbf{Satate} & \textbf{Worker1} & \textbf{Worker2} & \textbf{Worker3} \\
        \hline
        init & pod1 & pod2 & pod3, pod4 \\
        \hline
                1st violation (V1) & pod1 & pod2, pod3 & pod4  \\
        \hline
                2nd violation (V2) & pod1 & pod2, pod3 & pod4  \\
        \hline
                3rd violation (V3) & pod1 & pod2 & pod3, pod4  \\
        \hline
                4th violation (V4) & pod1 & pod2 & pod3, pod4  \\
        \hline
    \end{tabular}
    \captionsetup{justification=centering}
    \caption{Pod Placement State}
    \label{tab:podstate}
    \vspace{-0.2cm}
\end{table}

\insight{\toolname~effectively combines vertical and horizontal scaling with pod replacements to optimize performance, ensure stability, and maintain capacity constraints.}

\noindent \textbf{Comparison to Kubernetes Autoscaler: } We also conducted experiments using the Kubernetes Horizontal Pod Autoscaler (HPA)\footnote{https://kubernetes.io/docs/tasks/run-application/horizontal-pod-autoscale/} to demonstrate how our \toolname~outperforms the default scheduler in maintaining intent satisfaction. To the best of our knowledge, there is no comprehensive method that considers all aspects of the \toolname~together, making this comparison particularly valuable. The application was deployed with the same configuration described in the previous sections to ensure consistency and comparability. For the default scheduler, we configured a minimum of 1 replica for all pods and allowed scaling out to a maximum of 5 replicas as required. The CPU utilization thresholds for the autoscaler were set to 70\%, 60\%, and 50\% across three separate experiments. Note that identifying the best threshold for HPA to maintain application response time within a specific range can be application specific and challenging in practice. Thus, we varied thresholds to enable us to evaluate the autoscaler’s behavior in managing intent violations under different resource utilization conditions. 

As shown in Figure \ref{fig:70}, when the CPU threshold was set to 70\%, the autoscaler did not trigger any action to increase the number of replicas, even under increased loads. This was because the average CPU utilization never exceeded the threshold. However, when the CPU utilization threshold was reduced to 60\% and 50\%, Figure \ref{fig:60} and Figure \ref{fig:50} the autoscaler responded to resource demands by scaling up the number of replicas. Specifically, it increased the number of replicas for pod1 and pod3 which were under pressure based on the CPU threshold. 

\insight{The adaptive nature of \toolname~eliminates the complexity of threshold setting in the application auto-scaling process by focusing on high-level intents.}

\begin{figure}[t]
\centering
\includegraphics[width=0.95\columnwidth]{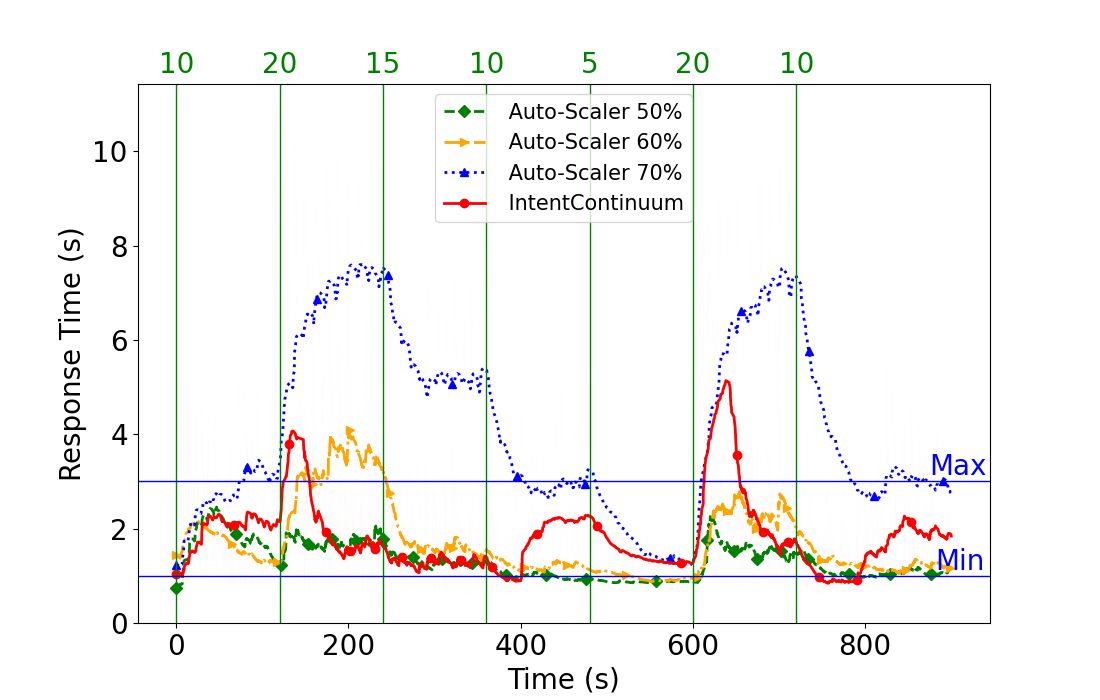}
\caption{Exponential moving average of the response time (EMA\_RT) Across Methods}
\label{Fig:app-rt-auto}
\end{figure}

Figure \ref{Fig:app-rt-auto} illustrates the effect of varying autoscaler thresholds on application response time, with \toolname~included for comparison. At a threshold of 70\%, the autoscaler does not scale up replicas, resulting in increased response times. In contrast, a threshold of 50\% often leads to an overprovisioning of replicas, as the number remains higher than necessary for much of the time. However, a threshold of 60\% strikes a balance by improving response time management and reducing resource utilization. Despite this improvement, the autoscaler is limited by its cooldown period, set to 5 minutes after detecting average CPU utilization, which prevents timely scaling down when the load decreases. Moreover, the autoscaler falls short in addressing violations as efficiently as \toolname~and struggles to manage upper and lower thresholds simultaneously with equal effectiveness, ultimately restricting precise control over resource scaling.
Table~\ref{tab:satisfy} presents the intent satisfaction rates and the total time during which the intent was violated for \toolname~and various configurations of the autoscaler. Based on the traffic patterns directed to the application, \toolname~consistently achieves a higher intent satisfaction rate compared to all autoscaler configurations. Specifically, \toolname~attains a satisfaction rate of approximately 85\%, outperforming the autoscaler thresholds of 70\%, 60\%, and 50\%, which achieve rates of 43\%, 79.5\%, and 82.5\%, respectively. This is also show in the time spent in violation compared to the autoscaler.

\begin{figure}[tph!t]
\centering
\includegraphics[width=0.9\columnwidth]{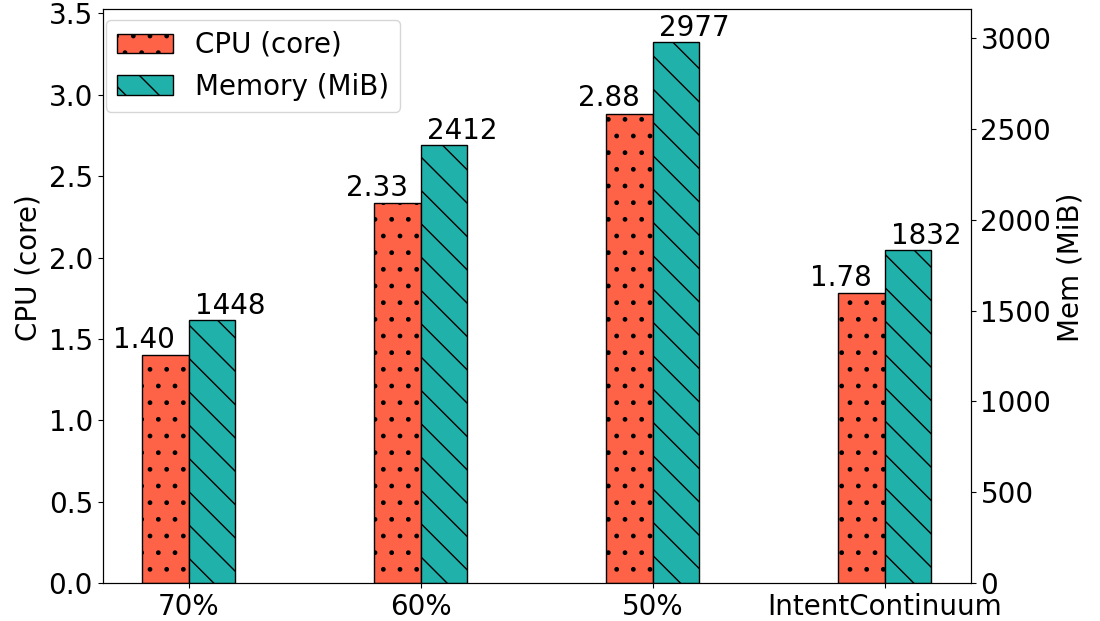}
\caption{Total Normalized Resource Usage Across Methods}
\label{Fig:cpu-mem}
\vspace{-0.2cm}
\end{figure}

\noindent \textbf{Resource Usage: }As shown in Table \ref{tab:normalized_all}, which reports normalized CPU and memory usage across all pods, \toolname~achieves the best resource utilization while adhering to the SLO defined in the intent. Similarly, Figure~\ref{Fig:cpu-mem} illustrates the combined normalized resource usage across the different methods. While the autoscaler configured with a 70\% threshold demonstrates the lowest CPU and memory usage, this comes at the cost of significantly higher application response times and the highest rate of intent violations. As the autoscaler threshold decreases, the system demonstrates increased resource usage and intent satisfaction, with the highest resource usage observed at a 50\% threshold. This is when the closest satisfaction rate to that of the \toolname~is achieved, but this approach requires approximately 60\% more CPU and memory usage. 

\insight{\toolname~provides the best balance between intent satisfaction and resource utilization compared to various Kubernetes autoscaler settings under varying traffic loads.}

\begin{table}[t]
\centering
\scriptsize
\renewcommand{\arraystretch}{1.2} % Adjust row height
\setlength{\tabcolsep}{5pt} % Adjust column width
\begin{tabular}{|p{3.4cm}|c|c|c|c|}
\hline
\multirow{2}{*}{\textbf{Metrics}} & \multirow{2}{*}{\textbf{\toolname~}} & \multicolumn{3}{c|}{\textbf{Autoscaler}} \\ \cline{3-5} 
                                  &                                           & \textbf{70\%} & \textbf{60\%} & \textbf{50\%} \\ \hline
Intent Satisfaction\%             & \textbf{85\%}                                      & 43\%          & 79.5\%        & 82.5\%        \\ \hline
%Number of Total \newline Requests          & 2824                                      & 1669          & 2859          & 3227          \\ \hline
Total Amount of Violated Time (s) & \textbf{143}                                       & 509           & 184           & 157           \\ \hline
\end{tabular}
\caption{Comparison of Metrics Across Methods}
\label{tab:satisfy}
\vspace{-0.6cm}
\end{table}

\begin{table}[h!]
    \centering
    \scriptsize
    \renewcommand{\arraystretch}{1.3} % Adjust row height
    \setlength{\tabcolsep}{6pt} % Adjust column spacing
    \begin{tabular}{|>{\centering\arraybackslash}m{1.5cm}|>{\centering\arraybackslash}m{1.5cm}|>{\centering\arraybackslash}m{0.8cm}|>{\centering\arraybackslash}m{0.8cm}|>{\centering\arraybackslash}m{0.8cm}|>{\centering\arraybackslash}m{0.8cm}|}
        \hline
        \textbf{Methods} & \textbf{Normalized} & \multicolumn{4}{c|}{\textbf{Pods}} \\
        \cline{3-6}
         & \textbf{Resource} & \textbf{P1} & \textbf{P2} & \textbf{P3} & \textbf{P4} \\
        \hline
        \multirow{2}{*}{\textbf{\shortstack{Intent \\ Continuum}}} & \shortstack{CPU (core)} & 0.515 & 0.3 & 0.665 & 0.3 \\
        \cline{2-6}
        & \shortstack{Mem (MiB)} & 530.46 & 312 & 677.195 & 312 \\
        \hline
        \multirow{2}{*}{\textbf{70\%}} & \shortstack{CPU (core)} & 0.3 & 0.3 & 0.5 & 0.3 \\
        \cline{2-6}
        & \shortstack{Mem (MiB)} & 312 & 312 & 512 & 312 \\
        \hline
        \multirow{2}{*}{\textbf{60\%}} & \shortstack{CPU (core)} & 0.79 & 0.3 & 0.95 & 0.3 \\
        \cline{2-6}
        & \shortstack{Mem (MiB)} & 818 & 312 & 970 & 312 \\
        \hline
        \multirow{2}{*}{\textbf{50\%}} & \shortstack{CPU (core)} & 1.04 & 0.3 & 1.24 & 0.3 \\
        \cline{2-6}
        & \shortstack{Mem (MiB)} & 1083 & 312 & 1270 & 312 \\
        \hline
    \end{tabular}
    \caption{Normalized Resource Utilization Across Methods}
\label{tab:normalized_all}
%\vspace{-0.6cm}
\end{table}

\subsubsection{\textbf{Networking Experiment}} Networking issues, such as link congestion, node failures, and link failures, can significantly affect application performance, making their consideration essential for robust intent management. Existing solutions in the literature often prioritize computing parameters while giving less emphasis to the combined impact of computing and networking factors.
In this section, we perform an experiment to illustrate that \toolname~effectively addresses intent violations, even those caused by networking issues. 

Locust is used to send HTTP requests to the application, with 10 users generating traffic at a spawn rate of 1 user per second over a duration of 900 seconds. Each request includes an image, and the generated traffic passes through the network switches as the pods are hosted on nodes interconnected through these switches. An initial traffic path is established between the pods, routed through the switches. Specifically, the initial route was configured as \textit{S2-S4-S3-S2}, determined by the placement of the pods on the respective nodes. The image size remained consistent with previous experiments, but unlike before, the user load in Locust is kept constant throughout the test because the focus is specifically on networking issues, such as link congestion. 

\begin{figure}[t]
\centering
\includegraphics[trim={0 0 0 35},clip, width=0.95\columnwidth]{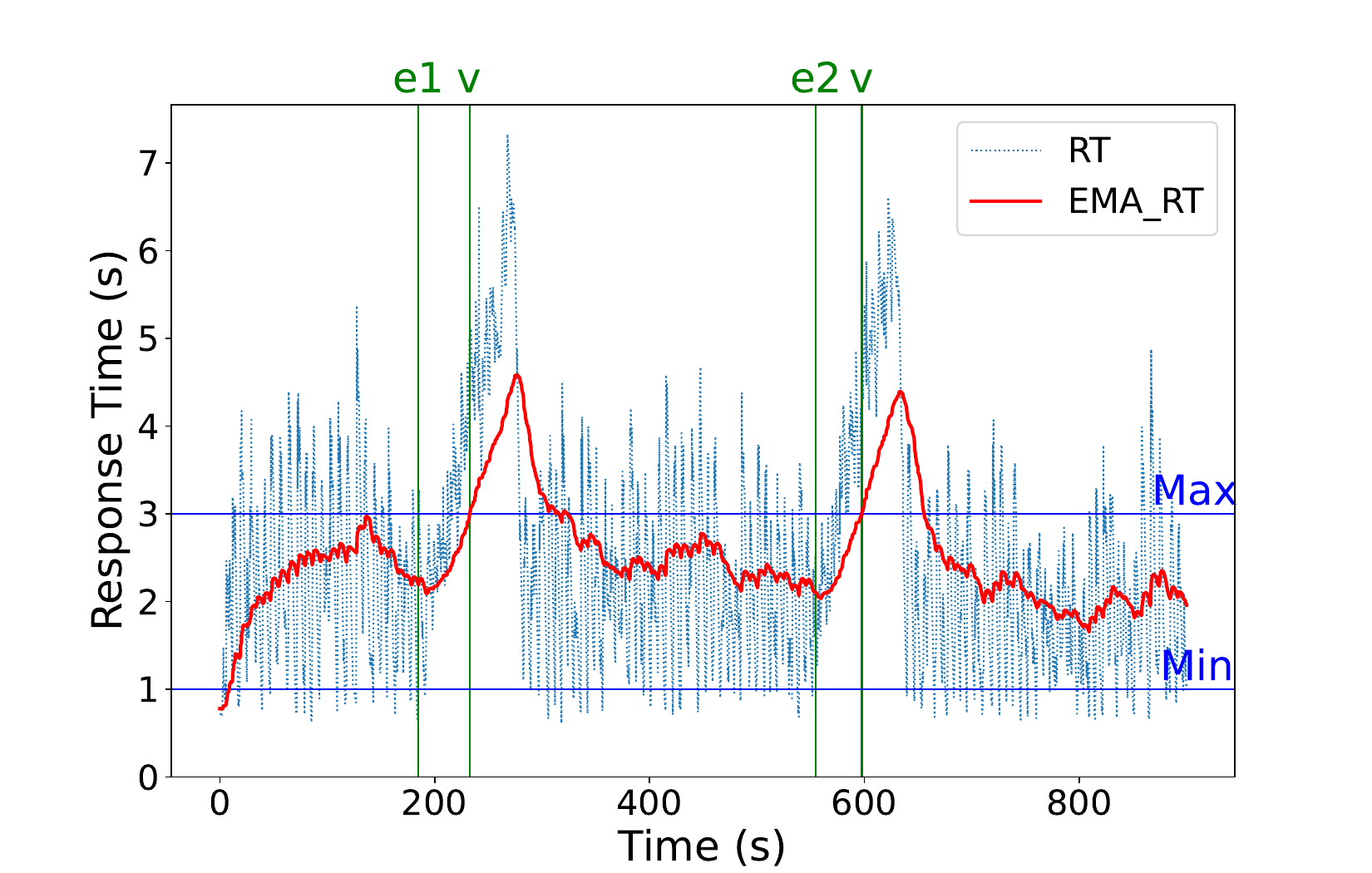}
\caption{Response Time for Networking Experiment}
\label{Fig:rt-flow}
\vspace{-0.6cm}
\end{figure}

Figure~\ref{Fig:rt-flow} illustrates the response time observed during this experiment. At $e1$, congestion occurs on the link between switches \textit{S2-S3}, introduced using \textit{iPerf}\footnote{https://iperf.fr/} with the help of two external hosts. As shown in the figure, a violation is detected after a delay (marked as v in the figure). \toolname~effectively mitigates the issue by recommending an alternative route to bypass the congested link. Initially, traffic was routed through \textit{S2-S4-S3-S2}; however, to address the congestion, the system implemented the recommended route via an API call to the SDN controller, which successfully updated the network flow. The new route utilized switches \textit{S2-S4-S3-S1-S2}, ensuring efficient traffic flow and resolving the violation. Later in the experiment, at $e2$, another instance of \textit{iPerf} was run simultaneously on two links, causing congestion between switches \textit{S1-S2} and \textit{S3-S4}. This congestion led to another violation, prompting \toolname~to send a request to the LLM asking for recommendations. The LLM suggested replacing pod3 from worker2 to worker1 to avoid the congested links. This action was carried out through an API call to the orchestrator, resulting in a new route being recommended: \textit{S2-S4-S2}.

\insight{\toolname~can address network issues such as link congestion or link failures by dynamically implementing flow scheduling or pod replacements through the combined use of an SDN controller and orchestrator.}

\insight{\toolname~effectively utilizes networking and computing metrics, either independently or in combination, to address intent violations.}

\subsection{Scalability Evaluation}\label{sec:scalable}
\begin{figure}[t]
\centering
\includegraphics[trim={0 0 0 35},clip, width=0.95\columnwidth]{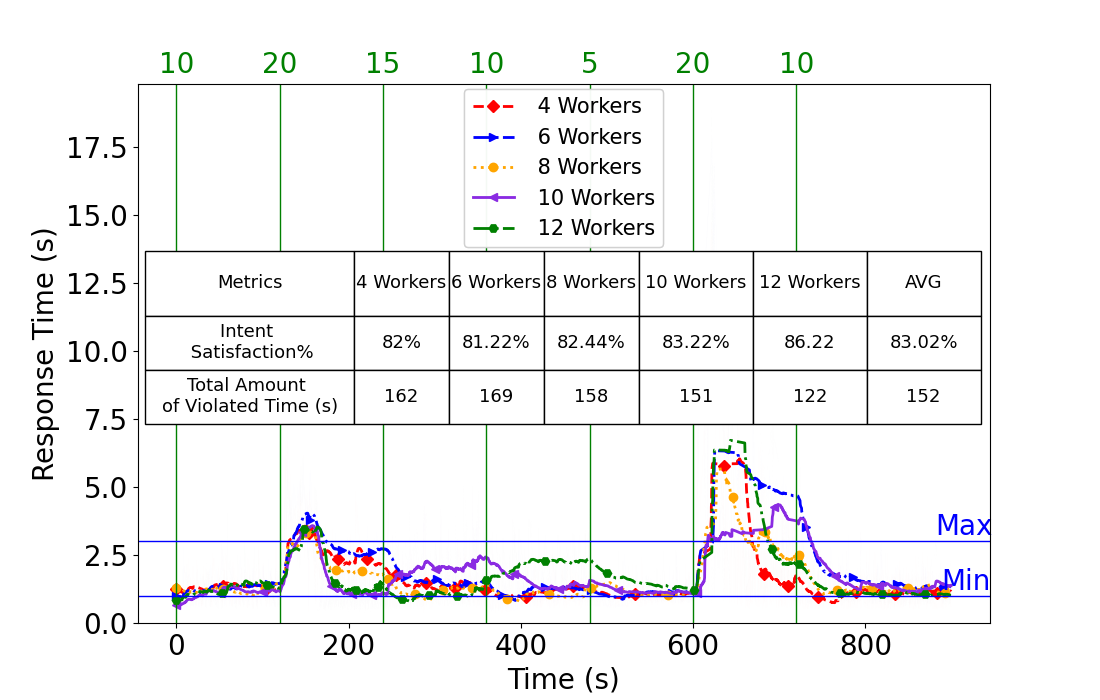}
\caption{EMA\_RT Across Senarios}
\label{Fig:table-nodes}
%\vspace{-0.6cm}
\end{figure}

In this section, we designed multiple scenarios with varying numbers of worker nodes to analyze the system's response to changes in node count. Given the constraints of our lab environment and the use of real cloud resources, we conducted experiments with 4, 6, 8, 10, and 12 worker nodes to evaluate the feasibility of \toolname~under different conditions. Figure \ref{Fig:table-nodes} presents the EMA\_RT across these scenarios, using the same traffic load as in Figure \ref{Fig:rtintent}. The results indicate that the intent satisfaction rate remains nearly consistent across different methods, with an average satisfaction rate of 83.02\%. To further evaluate scalability, we simulated scenarios with a larger number of nodes and sent the corresponding prompts to GPT-4o, measuring the latency—the time taken from when a prompt is sent to GPT-4o until a response is received. In each scenario, we introduced five violations and recorded both the average latency and the average token count sent to GPT-4o (tokens in) and received from it (tokens out) across these violations. This analysis illustrates how the system adapts as the number of nodes increases. The results are summarized in Table \ref{tab:gpt-token}. Notably, GPT-4o’s latency remained relatively stable across different scenarios, despite variations in the number of tokens processed. However, at 700 nodes, it failed to generate responses for some prompts due to token limitations, marking an upper boundary for direct scalability. Nevertheless, \toolname~can accommodate larger node counts by partitioning complex prompts into smaller, manageable segments, ensuring continued scalability. Further optimizations remain an area for future work.

\begin{comment}
\begin{table}[h]
    \centering
    \scriptsize
    \renewcommand{\arraystretch}{1.3} % Adjust row height
    \begin{tabular}{|p{1.7cm}|p{1.1cm}|p{1.2cm}|p{1.2cm}|p{1.2cm}|} % Adjust column widths here
        \hline
        \textbf{Metrics} & \textbf{3 Workers} & \textbf{4 Workers} & \textbf{5 Workers} & \textbf{6 Workers} \\
        \hline
        Intent Satisfaction\% & 84.11/\% & 77.33/\% & 78.5/\% & 63.77/\% \\
        \hline
        \makecell[l]{Total Amount \\ of Violation} & 143 & 204 & 193 & 326 \\
        \hline
    \end{tabular}
    \caption{Comparison of Metrics Across Scenarios}
    \label{tab:scale}
\end{table}
\end{comment}

% \begin{table}[h]
%     \centering
%     \scriptsize
%     \renewcommand{\arraystretch}{1.3} % Adjust row height
%     \begin{tabular}{|p{2.2cm}|p{2.2cm}|p{2.2cm}|} % Adjust column widths here
%         \hline
%         \textbf{Number of Worker Nodes} & \textbf{GPT-4o latency (s) AVG} & \textbf{Total Token Count AVG }  \\
%         \hline
%         10 & 13.37 & 5486  \\
%         \hline
%         50 & 13.95 & 7718  \\
%         \hline
%         100 & 22.10 & 10780  \\
%         \hline
%         200 & 14.50 & 16307  \\
%         \hline
%         300 & 12.80 & 21890  \\
%         \hline
%         400 & 12.69 & 27643  \\
%         \hline
%         500 & 12.45 & 33390  \\
%         \hline
%         600 & 13.42 & 39077  \\
%         \hline
%     \end{tabular}
%     \caption{Average GPT-4o Latency and Token Usage Across Different Node Scenarios (Over Five Violations)}
%     \label{tab:gpt-token}
% \end{table}

\begin{table}[h]
    \centering
    \scriptsize
    \setlength{\tabcolsep}{3pt} % Reduce column spacing (default is 6pt)
    \renewcommand{\arraystretch}{1.3} % Adjust row height
    \begin{tabular}{|p{1.6cm}|p{.65cm}|p{.65cm}|p{.65cm}|p{.65cm}|p{.65cm}|p{.65cm}|p{.65cm}|p{.65cm}|} % Fixed-width columns
        \hline
        \textbf{\# of Nodes} & 10 & 50 & 100 & 200 & 300 & 400 & 500 & 600 \\
        \hline
        \textbf{Latency (s)} & 13.37 & 13.95 & 13.80 & 14.50 & 12.80 & 12.69 & 12.45 & 13.42 \\
        \hline
        \textbf{\# of tokens in} & 4916 & 7196 & 10044 & 15745 & 21445 & 27145 & 32846 & 38552 \\
        \hline
        \textbf{\# of tokens out} & 570 & 522 & 736 & 561 & 446 & 498 & 544 & 525 \\
        \hline
    \end{tabular}
    \caption{Average GPT-4o Latency and Token Usage Across Different Node Scenarios over Five injected Violations}
    \label{tab:gpt-token}
    \vspace{-0.5cm}
\end{table}

\section{Discussions}\label{sec:discuss}
\subsection{Strengths}
Our \toolname~demonstrates several key strengths. It optimizes CPU and memory usage, efficiently manages replicas, and ensures that predefined intents and SLOs are consistently met within acceptable limits. By balancing resource allocation, it minimizes waste while adhering to defined thresholds. Additionally, \toolname~effectively manages both networking and computing parameters, demonstrating its versatility in handling diverse features. Note that while the framework leverages GPT-4o for decision-making, its design supports seamless integration with other LLMs. Moreover, LLM serves as auxiliary add-on features to the system, ensuring that even if it becomes unavailable, the application remains operational, albeit with potential impacts on QoS. In such cases, the framework can revert to built-in mechanisms for scheduling, autoscaling, and network flow management provided by tools like Kubernetes and the ONOS controller. This ensures effective handling of scenarios such as scaling, node failures, and network switch or link issues. This design enhances the robustness and resilience of the framework, ensuring it remains functional despite potential disruptions. 

\subsection{Limitations}
Despite its strengths, \toolname~has certain limitations that must be addressed to enhance its adaptability to a broader range of use cases and operational environments:

\noindent \textbf{Dependence on models:} The framework relies on GPT-4o for its decision-making. While this helps to improve automation, it may not always provide the best solutions in complex or unusual scenarios. The quality and accuracy of decisions also depend on the data fed into the model, which could affect performance in cases of incomplete or noisy data.

\noindent \textbf{Limited Transparency and Clarity of LLM Recommendations:} While GPT-4o can provide real-time resource management decisions, the reasoning behind some of its decisions may not be fully transparent. This black-box nature of LLMs could make it difficult to understand why certain actions were recommended, which may limit our \toolname~framework’s adoption in environments where transparency and clarity are critical. 

\noindent \textbf{Scalability and Processing Overhead:} Context Length for LLM models such as GPT-4o is limited. As the system scales up, the computational demands on GPT-4o could also introduce delays, particularly when processing large datasets in real time. In this paper, we focused on evaluating whether LLMs can provide effective recommendations for resource management in application deployments across the compute continuum. However, further research is required to explore their full scalability potential and identify possible performance bottlenecks at larger scales, which we leave as part of our future work.

\noindent \textbf{Financial implications:} As data volumes grow, the number of tokens exchanged with the LLM increases, resulting in higher costs. This impact becomes more pronounced as the application scales, incorporating a larger number of edge nodes, cloud nodes, IoT devices, and microservices.

\section{Related Work}\label{sec:related}
Research on resource management in the compute continuum has explored various techniques for optimizing task allocation and system performance. These techniques range from traditional algorithms to emerging AI-driven methods, each with unique strengths and limitations.

\begin{table*}[htbp]
\scriptsize
\centering
\renewcommand{\arraystretch}{1} % Adjust row height
\begin{tabular}{|>{\raggedright\arraybackslash}p{1.5cm}|>{\raggedright\arraybackslash}p{2.9cm}|>{\raggedright\arraybackslash}p{4.9cm}|>{\raggedright\arraybackslash}p{7cm}|} % Adjust column width
\hline
\textbf{Approach} & \textbf{Key Parameters} & \textbf{Strengths} & \textbf{Gaps} \\ \hline

Traditional Resource Management & Heuristics, meta-heuristics (GA, PSO, Simulated Annealing)  & Scalable, near-optimal solutions for smaller systems, low computational complexity  & Lacks real-time adaptability, requires manual tuning, inefficient for dynamic and unpredictable environments \\ \hline

Reinforcement Learning (AI-Driven) & State-action feedback, reward functions, energy and latency optimization & Learns from system feedback, dynamically adjusts resource allocation, improves energy efficiency and latency & Requires large training data, struggles to generalize across different systems, focused on specific performance parameters (e.g., energy, latency) \\ \hline

Deep Learning Models (AI-Driven) & Historical data, workload prediction, system forecasting  & Predicts future resource needs based on historical patterns, enables pre-emptive resource allocation & Lacks adaptability to sudden or unforeseen changes, ineffective for real-time decision-making \\ \hline

LLM (Large Language Models) & Contextual reasoning, real-time data analysis, system monitoring  & Provides context-aware decision-making, processes large amounts of data, adaptable to dynamic environments & Limited research on real-time applications, underexplored integration with intent-driven frameworks, requires further development for practical deployment in compute continuum environments \\ \hline

\end{tabular}
\caption{Resource Management Approaches and Gaps}
\label{tab:oldmethod}
\vspace{-0.4cm}
\end{table*}

\subsection{Traditional Resource Management Approaches}
Traditional resource allocation methods, such as heuristic and meta-heuristic algorithms, are widely used for their ability to provide near-optimal solutions efficiently \cite{sharma2022systematic}. Techniques like genetic algorithms (GA) \cite{katoch2021review}, particle swarm optimization (PSO) \cite{shami2022particle}, and simulated annealing \cite{guilmeau2021simulated} excel in static or smaller systems but struggle with the scalability and real-time adaptability required in dynamic compute continuum environments \cite{xhafa2008metaheuristics}. While advancements like dynamic parameter tuning in GA \cite{materwala2023qos} improve flexibility, these methods often rely on manual adjustments, limiting their effectiveness in unpredictable, distributed systems \cite{10053388}.

\subsection{Edge-Cloud Resource Management and AI-Driven Intents}
Resource allocation across edge and cloud infrastructures remains a critical research area, with strategies like task offloading reducing latency and energy consumption \cite{liu2019energy}. However, static policies often fail under rapidly changing workloads, prompting the adoption of dynamic workload partitioning \cite{cao2020dynamic}. While these advances improve adaptability, the real-time responsiveness required for complex IoT systems remains challenging, as current methods struggle to continuously monitor and respond to environmental changes.

AI techniques, including reinforcement learning (RL) and predictive deep learning models, have enhanced resource management by enabling automated adjustments and anticipating resource needs \cite{zheng2022deep}. RL has been effective in optimizing specific parameters like latency \cite{khani2024deep} but requires extensive training data, limiting its applicability to dynamic environments \cite{wang2024optimizing}. Predictive models improve efficiency using historical data but often fail to adapt to unforeseen changes \cite{hurtado2023continual}.

Intent-driven systems offer a promising approach, translating user-defined goals into automated policies for resource management \cite{kyryk2021infrastructure}. While these systems provide greater flexibility than traditional methods, their reliance on static intent mappings limits their adaptability to the dynamic nature of edge-cloud environments \cite{sicari2023toward}. Future systems must incorporate more dynamic, context-aware mechanisms to address the challenges of distributed, heterogeneous environments.

\subsection{Large Language Models in System Management}
Large language models like ChatGPT and Llama are transforming system management in distributed environments such as the compute continuum. With their ability to process natural language, these models automate decision-making, detect anomalies, predict resource bottlenecks, and optimize scheduling and resource allocation \cite{mongaillard2024large, ji2022intelligent}. Their adaptability to changing conditions enables proactive, intent-driven management, reducing human intervention and improving operational efficiency.

\subsection{Gap in Existing Research}
Table \ref{tab:oldmethod} compares existing resource management approaches, highlighting key parameters and gaps. Traditional methods, though scalable, struggle with real-time dynamics due to manual configurations. AI-driven methods improve adaptability but need large datasets and lack holistic context understanding. LLMs show promise for context-aware decisions but are underexplored in real-time, intent-based management.

\section{Conclusions and Future Work}\label{sec:sumup}
In this paper, we presented a comprehensive framework for intent-driven resource management in the compute continuum, leveraging the capabilities of LLMs to address the complexities associated with deploying applications across edge and cloud environments. The proposed \toolname~framework effectively ensures that user-defined intents, such as maintaining application response times within a specified range, are consistently satisfied. By incorporating real-time system monitoring, root cause analysis, and adaptive corrective actions, \toolname~significantly reduces the need for human intervention and simplifies the complexity of resource management.
Furthermore, \toolname~demonstrates the ability to manage both networking and computing parameters simultaneously, ensuring seamless operation and improved system reliability. Our approach outperforms traditional methods by maintaining system stability under varying workloads, optimizing resource utilization, and dynamically adapting to changing conditions. The findings from our real-world proof-of-concept implementation and experimental evaluation underscore the effectiveness of our proposed method.

For future work, we aim to enhance the transparency of the framework, reduce computational overhead, and further improve its cost-effectiveness and scalability, enabling broader adoption and applicability in real-world scenarios.

\bibliographystyle{IEEEtran}

\begin{thebibliography}{10}
\providecommand{\url}[1]{#1}
\csname url@samestyle\endcsname
\providecommand{\newblock}{\relax}
\providecommand{\bibinfo}[2]{#2}
\providecommand{\BIBentrySTDinterwordspacing}{\spaceskip=0pt\relax}
\providecommand{\BIBentryALTinterwordstretchfactor}{4}
\providecommand{\BIBentryALTinterwordspacing}{\spaceskip=\fontdimen2\font plus
\BIBentryALTinterwordstretchfactor\fontdimen3\font minus \fontdimen4\font\relax}
\providecommand{\BIBforeignlanguage}[2]{{%
\expandafter\ifx\csname l@#1\endcsname\relax
\typeout{** WARNING: IEEEtran.bst: No hyphenation pattern has been}%
\typeout{** loaded for the language `#1'. Using the pattern for}%
\typeout{** the default language instead.}%
\else
\language=\csname l@#1\endcsname
\fi
#2}}
\providecommand{\BIBdecl}{\relax}
\BIBdecl

\bibitem{lee2015internet}
I.~Lee and K.~Lee, ``The internet of things (iot): Applications, investments, and challenges for enterprises,'' \emph{Business horizons}, vol.~58, no.~4, pp. 431--440, 2015.

\bibitem{tabrizchi2020survey}
H.~Tabrizchi and M.~Kuchaki~Rafsanjani, ``A survey on security challenges in cloud computing: issues, threats, and solutions,'' \emph{The journal of supercomputing}, vol.~76, no.~12, pp. 9493--9532, 2020.

\bibitem{russo2023serverless}
G.~R. Russo, V.~Cardellini, and F.~L. Presti, ``Serverless functions in the cloud-edge continuum: Challenges and opportunities,'' in \emph{2023 31st Euromicro International Conference on Parallel, Distributed and Network-Based Processing (PDP)}.\hskip 1em plus 0.5em minus 0.4em\relax IEEE, 2023, pp. 321--328.

\bibitem{9083958}
K.~Cao, Y.~Liu, G.~Meng, and Q.~Sun, ``An overview on edge computing research,'' \emph{IEEE Access}, vol.~8, pp. 85\,714--85\,728, 2020.

\bibitem{sajid2013cloud}
M.~Sajid and Z.~Raza, ``Cloud computing: Issues \& challenges,'' in \emph{International conference on cloud, big data and trust}, vol.~20, no.~13.\hskip 1em plus 0.5em minus 0.4em\relax sn, 2013, pp. 13--15.

\bibitem{danelutto2024structuring}
M.~Danelutto, P.~Dazzi, and M.~Torquati, ``Structuring the continuum,'' in \emph{International Conference on Advanced Information Networking and Applications}.\hskip 1em plus 0.5em minus 0.4em\relax Springer, 2024, pp. 212--223.

\bibitem{guzek2015survey}
M.~Guzek, P.~Bouvry, and E.-G. Talbi, ``A survey of evolutionary computation for resource management of processing in cloud computing,'' \emph{IEEE Computational Intelligence Magazine}, vol.~10, no.~2, pp. 53--67, 2015.

\bibitem{sangaiah2020iot}
A.~K. Sangaiah, A.~A.~R. Hosseinabadi, M.~B. Shareh, S.~Y. Bozorgi~Rad, A.~Zolfagharian, and N.~Chilamkurti, ``Iot resource allocation and optimization based on heuristic algorithm,'' \emph{Sensors}, vol.~20, no.~2, p. 539, 2020.

\bibitem{sharma2022systematic}
V.~Sharma and A.~K. Tripathi, ``A systematic review of meta-heuristic algorithms in iot based application,'' \emph{Array}, vol.~14, p. 100164, 2022.

\bibitem{katoch2021review}
S.~Katoch, S.~S. Chauhan, and V.~Kumar, ``A review on genetic algorithm: past, present, and future,'' \emph{Multimedia tools and applications}, vol.~80, pp. 8091--8126, 2021.

\bibitem{guilmeau2021simulated}
T.~Guilmeau, E.~Chouzenoux, and V.~Elvira, ``Simulated annealing: A review and a new scheme,'' in \emph{2021 IEEE statistical signal processing workshop (SSP)}.\hskip 1em plus 0.5em minus 0.4em\relax IEEE, 2021, pp. 101--105.

\bibitem{millnert2020holoscale}
V.~Millnert and J.~Eker, ``Holoscale: Horizontal and vertical scaling of cloud resources,'' in \emph{2020 IEEE/ACM 13th International Conference on Utility and Cloud Computing (UCC)}.\hskip 1em plus 0.5em minus 0.4em\relax IEEE, 2020, pp. 196--205.

\bibitem{marchese2022network}
A.~Marchese and O.~Tomarchio, ``Network-aware container placement in cloud-edge kubernetes clusters,'' in \emph{2022 22nd IEEE international symposium on cluster, cloud and internet computing (CCGrid)}.\hskip 1em plus 0.5em minus 0.4em\relax IEEE, 2022, pp. 859--865.

\bibitem{toosi2019elasticsfc}
A.~N. Toosi, J.~Son, Q.~Chi, and R.~Buyya, ``Elasticsfc: Auto-scaling techniques for elastic service function chaining in network functions virtualization-based clouds,'' \emph{Journal of Systems and Software}, vol. 152, pp. 108--119, 2019.

\bibitem{10643932}
N.~Akbari, A.~N. Toosi, J.~Grundy, H.~Khalajzadeh, M.~S. Aslanpour, and S.~Ilager, ``icontinuum: An emulation toolkit for intent-based computing across the edge-to-cloud continuum,'' in \emph{2024 IEEE 17th International Conference on Cloud Computing (CLOUD)}, 2024, pp. 468--474.

\bibitem{fareghzadeh2018dynamic}
N.~Fareghzadeh, M.~A. Seyyedi, and M.~Mohsenzadeh, ``Dynamic performance isolation management for cloud computing services,'' \emph{The Journal of Supercomputing}, vol.~74, pp. 417--455, 2018.

\bibitem{brown2020language}
T.~Brown, B.~Mann, N.~Ryder, M.~Subbiah, J.~D. Kaplan, P.~Dhariwal, A.~Neelakantan, P.~Shyam, G.~Sastry, A.~Askell \emph{et~al.}, ``Language models are few-shot learners,'' \emph{Advances in neural information processing systems}, vol.~33, pp. 1877--1901, 2020.

\bibitem{shami2022particle}
T.~M. Shami, A.~A. El-Saleh, M.~Alswaitti, Q.~Al-Tashi, M.~A. Summakieh, and S.~Mirjalili, ``Particle swarm optimization: A comprehensive survey,'' \emph{Ieee Access}, vol.~10, pp. 10\,031--10\,061, 2022.

\bibitem{xhafa2008metaheuristics}
F.~Xhafa and A.~Abraham, \emph{Metaheuristics for scheduling in distributed computing environments}.\hskip 1em plus 0.5em minus 0.4em\relax Springer, 2008, vol. 146.

\bibitem{materwala2023qos}
H.~Materwala, L.~Ismail, and H.~S. Hassanein, ``Qos-sla-aware adaptive genetic algorithm for multi-request offloading in integrated edge-cloud computing in internet of vehicles,'' \emph{Vehicular Communications}, vol.~43, p. 100654, 2023.

\bibitem{10053388}
S.~Gupta and N.~Singh, ``Heuristics and meta-heuristics based algorithms for resource optimization in fog computing environment: A comparative study,'' in \emph{2023 International Conference on Intelligent Data Communication Technologies and Internet of Things (IDCIoT)}, 2023, pp. 271--276.

\bibitem{liu2019energy}
F.~Liu, Z.~Huang, and L.~Wang, ``Energy-efficient collaborative task computation offloading in cloud-assisted edge computing for iot sensors,'' \emph{Sensors}, vol.~19, no.~5, p. 1105, 2019.

\bibitem{cao2020dynamic}
Z.~Cao, B.~Xiao, H.~Duan, L.~Yang, and W.~Cai, ``A dynamic partitioning framework for edge-assisted cloud computing,'' in \emph{International Conference on Algorithms and Architectures for Parallel Processing}.\hskip 1em plus 0.5em minus 0.4em\relax Springer, 2020, pp. 215--229.

\bibitem{zheng2022deep}
T.~Zheng, J.~Wan, J.~Zhang, and C.~Jiang, ``Deep reinforcement learning-based workload scheduling for edge computing,'' \emph{Journal of Cloud Computing}, vol.~11, no.~1, p.~3, 2022.

\bibitem{khani2024deep}
M.~Khani, M.~M. Sadr, and S.~Jamali, ``Deep reinforcement learning-based resource allocation in multi-access edge computing,'' \emph{Concurrency and Computation: Practice and Experience}, vol.~36, no.~15, p. e7995, 2024.

\bibitem{wang2024optimizing}
S.~Wang, Y.~Li, and F.~Chen, ``Optimizing blue team strategies with reinforcement learning for enhanced ransomware defense simulations,'' 2024.

\bibitem{hurtado2023continual}
J.~Hurtado, D.~Salvati, R.~Semola, M.~Bosio, and V.~Lomonaco, ``Continual learning for predictive maintenance: Overview and challenges,'' \emph{Intelligent Systems with Applications}, vol.~19, p. 200251, 2023.

\bibitem{kyryk2021infrastructure}
M.~Kyryk, N.~Pleskanka, M.~Pleskanka, and V.~Kyryk, ``Infrastructure as code and microservices for intent-based cloud networking,'' in \emph{Future Intent-Based Networking: On the QoS Robust and Energy Efficient Heterogeneous Software Defined Networks}.\hskip 1em plus 0.5em minus 0.4em\relax Springer, 2021, pp. 51--68.

\bibitem{sicari2023toward}
C.~Sicari, A.~Catalfamo, L.~Carnevale, A.~Galletta, A.~Celesti, M.~Fazio, and M.~Villari, ``Toward the edge-cloud continuum through the serverless workflows,'' in \emph{Device-Edge-Cloud Continuum: Paradigms, Architectures and Applications}.\hskip 1em plus 0.5em minus 0.4em\relax Springer, 2023, pp. 1--18.

\bibitem{mongaillard2024large}
T.~Mongaillard, S.~Lasaulce, O.~Hicheur, C.~Zhang, L.~Bariah, V.~S. Varma, H.~Zou, Q.~Zhao, and M.~Debbah, ``Large language models for power scheduling: A user-centric approach,'' \emph{arXiv preprint arXiv:2407.00476}, 2024.

\bibitem{ji2022intelligent}
Z.~Ji, J.~Zhang, and X.~Wang, ``Intelligent scheduling strategies for computing power resources in heterogeneous edge networks,'' in \emph{International Conference of Pioneering Computer Scientists, Engineers and Educators}.\hskip 1em plus 0.5em minus 0.4em\relax Springer, 2022, pp. 253--271.

\end{thebibliography}
  % Generated by IEEEtran.bst, version: 1.14 (2015/08/26)

% that's all folks
\end{document}